\documentclass[pre,preprint,showpacs,preprintnumbers,amsmath,amssymb,nofootinbib,showkeys]{revtex4}

\usepackage{graphicx}

\usepackage{dcolumn}

\usepackage{bm}

\begin{document}

\preprint{APS/123-QED}

\title{The origin of chaos near critical points of the quantum flow}

\author{C. Efthymiopoulos}
 \email{cefthim@academyofathens.gr}

 \author{C. Kalapotharakos}
 \email{ckalapot@phys.uoa.gr}

\author{G. Contopoulos}
 \email{gcontop@academyofathens.gr}

\affiliation{
      Research Center for Astronomy and Applied Mathematics, Academy of Athens,
      Soranou Efesiou 4, GR-115 27 Athens, Greece
           }

\begin{abstract}

The general theory of motion in the vicinity of a moving quantum
nodal point (vortex) is studied in the framework of the de Broglie - Bohm
trajectory method of quantum mechanics. Using an adiabatic
approximation, we find that near any nodal point of an arbitrary
wavefunction $\psi$ there is an unstable point (called X-point)
in a frame of reference moving with the nodal point. The local
phase portrait forms always a characteristic pattern called the
`nodal point - X-point complex'. We find general formulae for this
complex as well as necessary and sufficient conditions of validity
of the adiabatic approximation. We demonstrate that chaos
emerges from the consecutive
scattering events of the orbits with nodal point - X-point complexes.
The scattering events are of two types (called type I and
type II). A theoretical model is constructed yielding the local
value of the Lyapunov characteristic number in scattering events of both
types. The local Lyapunov characteristic number  scales as an inverse power of the
speed of the nodal point in the rest frame, implying that it
scales proportionally to the size of the nodal point X- point
complex. It is also an inverse power of the distance of a
trajectory from the X-point's stable manifold far from the
complex. This distance plays the role of an effective `impact
parameter'. The results of detailed numerical experiments with
different wavefunctions, possessing one, two, or three moving
nodal points, are reported. Examples are given of regular and
chaotic trajectories, and the statistics of the Lyapunov characteristic
number of the orbits are found and compared to the number of encounter
events of each orbit with the nodal point X-point complexes. The
numerical results are in agreement with the theory, and various
phenomena appearing at first as counter-intuitive find a
straightforward explanation.

\end{abstract}


\maketitle

\section{Introduction}

The dynamics in quantum systems with {\it vortices}, i.e.
singularities of the phase field of the wavefunction
$\psi=Re^{iS/\hbar}$ \cite{dirac1931,hgb1974}, has attracted much
interest in recent years because of a variety of potential
applications, e.g. in tunneling through potential barriers,
\cite{hcp1974,srv1989,lw1999,bwf2003}, ballistic electron
transport \cite{bvh1991,ws1994a,bss2001}, superfluidity
\cite{feynman1955}, Bose-Einstein condensates
\cite{ds1996,rok1997,sf1998,dz1999,grpg1999}, optical lattices
\cite{vft2007}, atom-surface scattering \cite{sbm2004}, Josephson
junctions \cite{bglmo1999}, decoherence \cite{nw2002} etc. The
so-called `trajectory' approach lends itself very conveniently to
such a study  (see \cite{w2005} for a review). In this approach
one follows the orbits of `particles' tracing the
quantum-mechanical currents. This is computationally equivalent to
a Lagrangian quantum-hydrodynamical approach \cite{madelung1926}
or to the `Bohmian' or `pilot wave' approach
\cite{brog1926,bohm1952a,bohm1952b}. The trajectories are
described by first order equations of the form ${\mathbf v}=\nabla
S(\mathbf{x})/\hbar$. Such trajectories provide a Lagrangian
visualization of quantum processes (e.g. \cite{hol2005}) which is
distinct from the Eulerian (i.e. Schr\"{o}dinger) approach,
although it is consistent with it.

A number of authors have found that the quantum trajectories of
low-dimensional systems can be very {\it chaotic}
\cite{dgz1992,fs1995,pv1995,pol1996,dm1996,ip1996,frisk1997,km1998,
ws1999,mpd2000,cush2000,ff2003,sf2003,wp2005,vw2005,wpb2007,sf2008}.
The physical importance of chaotic quantum trajectories has been
extensively discussed in three recent papers of ours
\cite{efthcont2006,ekc2007,contefth2008}.

The generation of chaos is directly associated with the appearance
of quantum vortices. It has been pointed out \cite{wp2005,wpb2007}
that chaos is, in general, caused by the {\it motion} of quantum
vortices. In the case of fixed vortices, on the other hand, chaos
can still be generated if we allow a coupling of the wavefunction
to a vector electromagnetic potential (e.g. \cite{ws1994b,ws1999}).

In previous papers of ours (\cite{ekc2007}, hereafter EKC, and
\cite{contefth2008}) we studied the quantum trajectories in particular
examples of 2D systems with a moving quantum vortex, in order to find
numerical indications of the mechanism of generation of chaos. We note that
this problem is quite different from traditional problems of nonlinear
dynamics. First, the equations of motion become singular near a vortex.
Furthermore, the vortex is oscillating quasi-periodically, i.e., with
more than one incommensurate frequency. Thus there are no obviously
identifiable critical points of the flow other than the vortex itself.

Setting the identification of the critical points as a primary target,
in EKC we looked for such points in a {\it moving local frame of
reference} centered at a moving nodal point, and made use of an
approximative form of the equations of motion valid under a so-called
adiabatic approximation. Our main finding can be summarized as follows:
In the above frame and approximation, the nodal point is seen to create
a saddle point nearby, called the `X-point'. The local phase portrait
was called the `nodal point - X-point complex'. Most trajectories do not
penetrate deeply inside the complex. Instead, they are accelerated along
the X-point's asymptotic curves and eventually they are scattered by the
complex. Chaos is generated by  a sequence of such scattering events.
This conclusion was substantiated by following the evolution of the
deviation vectors of some representative chaotic trajectories. We also
found the domains avoided by the nodal point - X point complex and
demonstrated that the trajectories covering these domains are regular,
i.e. they obey effective integrals of motion and they have zero Lyapunov
characteristic numbers.

Our study so far relied only on numerical examples of trajectories
in particular $\psi-$fields, in which (i) the wavefunction $\psi(x,y,t)$
has a simple form, and (ii) there is only one nodal point present in the
configuration space $(x,y)$ at any time $t$. These restrictions are
removed in the present paper, in which (i) we develop the general theory
of motion near moving 2D quantum vortices, applicable to generic $\psi-$
fields, and (ii) we continue the study of particular numerical examples,
in cases with more than one nodal point.

Regarding (i), section II contains a general analytical treatment of the
motion near the critical points of the quantum flow yielding: a) the form
of the equations of motion in terms of the coefficients of a local expansion
of a generic $\psi-$field around a nodal point, b) general formulae for the
structure and stability of the nodal point - X-point complex, c) conditions
of validity of the adiabatic approximation, and (most importantly) d) theoretical
predictions for the values of the Lyapunov characteristic numbers generated locally
by the interaction of the trajectories with a nodal point - X-point complex.
The latter problem is treated like a classical scattering problem. The most
important parameters in the theory turn to be the speed of the nodal point
and the impact parameter, i.e. distance of a trajectory from the X-point's
stable manifold far from the complex. The theory leads to a quantification
of the degree of chaos, i.e. the level of the Lyapunov characteristic number
of the trajectories, in systems with quantum vortices.

Regarding (ii), Section III tests the theory of section II against numerical
experiments, focusing on examples in which the $\psi-$field generates more
than one nodal point at the same time. We obtain numerical values of
the Lyapunov characteristic numbers for statistical ensembles of orbits and compare
these values with the predictions of section II. We also check the quantitative
relation between the Lyapunov characteristic number and the number of encounters of a
trajectory with the nodal point - X-point complexes. Section IV summarizes
our conclusions.

\section{The motion in the vicinity of a nodal point}

\subsection{Equations of motion}

Let $\bigg(x_0(t),y_0(t)\bigg)$ represent the center of a moving
frame of reference,
$\vec{V}(t)\equiv(V_x,V_y)=(\dot{x}_0,\dot{y}_0)$ being its
velocity at the time $t$ with respect to the rest frame $(x,y)$.
The wavefunction can be expanded around $(x_0,y_0)$. Up to second
degree we have:
\begin{eqnarray}\label{psiexp}
\psi&=&\psi_0(t)+\bigg(a_{10}(t)+ib_{10}(t)\bigg)u+\bigg(a_{01}(t)+ib_{01}(t)\bigg)v
+{1\over
2}\bigg(a_{20}(t)+ib_{20}(t)\bigg)u^2\nonumber\\
&+&{1\over 2}\bigg(a_{02}(t)+ib_{02}(t)\bigg)v^2
+\bigg(a_{11}(t)+ib_{11}(t)\bigg)u v+\ldots
\end{eqnarray}
where $u=x-x_0$, $v=y-y_0$ and the coefficients $a_{ij}$, $b_{ij}$
are real.

The equations of motion in such a frame, which follow from the
equations of motion $\dot{\vec{x}}=\nabla S$ ($\hbar=1$) in the
rest frame, read:
\begin{equation}\label{eqmouv}
(\dot{u},\dot{v})=Im\bigg({\nabla_{u,v}\psi\over\psi}\bigg)
-(V_x,V_y)~~.
\end{equation}

All the frames of reference moving with the same velocities
$V_x(t),V_y(t)$ at all times $t$ form an equivalence class with
respect to parallel translations in the configuration space. We
consider as representative of the class a frame, of which the
center $\bigg(x_0(t_0),y_0(t_0)\bigg)$ coincides at some time
$t=t_0$ with the instantaneous position of a nodal point
$(x_N(t_0),y_N(t_0))$ of the wavefunction $\psi$, i.e.
$x_0(t_0)=x_N(t_0)$, $y_0(t_0)=y_N(t_0)$. Assume further that
$(x_N,y_N)$ is a simple root of the system of equations
$Re(\psi)=Im(\psi)=0$. Eq.(\ref{psiexp}) implies that
$\psi_0(t_0)=0$, but not all the coefficients $a_{10}$, $a_{01}$,
$b_{10}$, $b_{01}$ vanish at $t=t_0$. A third requirement is that the
field current
$j=[Re(\psi)\nabla(Im(\psi))-Im(\psi)\nabla(Re(\psi))]/(2i)$ should be
divergence-free, $\nabla\cdot j=0$, at the position of the nodal
point. This follows from the continuity equation
$\partial\rho/\partial t + \nabla\cdot j=0$, since any zero of the
wavefunction $\psi=0$ is a local spatio-temporal minimum of
$\rho=|\psi|^2$, thus $\partial\rho/\partial t = 0$ at
$(x_0(t_0),y_0(t_0))$. The condition $\nabla\cdot j=0$ implies
that
\begin{equation}\label{curfree}
a_{02}=-a_{20},~~b_{02}=-b_{20}~~.
\end{equation}

Substituting the expansion $(\ref{psiexp})$ into (\ref{eqmouv}),
for $t=t_0$, and taking into account the previous conditions, the
equations of motion take the form (up to second degree):
\begin{eqnarray}\label{eqmoexp}
{du\over dt}&=&{1\over G}\times\bigg[ (a_{01}b_{10}-a_{10}b_{01})v
+ {1\over 2}(a_{02}b_{10}-a_{10}b_{02})u^2 + ({1\over
2}a_{02}b_{10}-{1\over 2}a_{10}b_{02}-a_{11}b_{01})v^2 \nonumber\\
& &+ (a_{02}b_{01}-a_{01}b_{02})u~v+\ldots\bigg]-V_x\\
{dv\over dt}&=&{1\over G}\times\bigg[ (a_{10}b_{01}-a_{01}b_{10})u
+ {1\over 2}(a_{01}b_{02}-a_{02}b_{01})v^2 + ({1\over
2}a_{01}b_{02}-{1\over 2}a_{02}b_{01}-a_{11}b_{10})u^2\nonumber \\
& &+ (a_{10}b_{02}-a_{02}b_{10})u~v+\ldots\bigg]-V_y\nonumber
\end{eqnarray}
with
\begin{equation}\label{gi}
G=(a_{10}^2+b_{10}^2)u^2+(a_{01}^2+b_{01}^2)v^2+2(a_{01}a_{10}+b_{01}b_{10})u~v+\ldots
\end{equation}

Equations (\ref{eqmoexp}) yield the ensemble of instantaneous flow
lines (phase portrait) in the selected frame of reference for $t=t_0$.
Two questions are now examined, namely a) the typical form of the
instantaneous phase portrait, and b) whether {\it adiabatic conditions}
are satisfied, ensuring that the form of the phase portrait changes in
time slowly, relative to the typical velocities along the particles'
trajectories within this portrait.

\subsection{Phase portrait: nodal point - X-point complex}

The adiabatic approximation for $t=t_0$ holds in space domains in
which $du/dt$, $dv/dt$ are large compared to the time derivatives
of the coefficients $a_{ij}$, $b_{ij}$, and of the velocities
$V_x,V_y$. We can then `freeze' $a_{ij}$, $b_{ij}$, $V_x$, $V_y$
to their fixed values at $t=t_0$ and treat the flow
(\ref{eqmoexp}) as autonomous. In such an approximation we find
the following features of the instantaneous phase portrait around
the nodal point $(u,v)=(0,0)$:

\subsubsection{Nodal point}

In polar coordinates $u=R\cos\phi$, $v=R\sin\phi$ Eqs.(\ref{eqmoexp})
take the form:
\begin{equation}\label{eqmo2}
{dR\over dt}={c_2R^2+c_3R^3+c_4R^4+...\over G},~~~ {d\phi\over
dt}={d_0+d_1R+d_2R^2+...\over G}
\end{equation}
where the coefficients $c_j$ and $d_j$ depend on a) the
coefficients $a_{ij}$, $b_{ij}$, b) the velocities $(V_x,V_y)$,
and c) powers of the trigonometric functions $\sin\phi,\cos\phi$.
The lowest order terms of $G$, given by Eq.(\ref{gi}), read:
\begin{equation}\label{giexp}
G=R^2\bigg[(a_{10}^2+b_{10}^2)\cos^2\phi+(a_{01}^2+b_{01}^2)\sin^2\phi
+(a_{01}a_{10}+b_{01}b_{10})\sin 2\phi\bigg]+O(R^3)~~.
\end{equation}
The quantity in the square bracket of (\ref{giexp}) is always positive.
Thus, the second of Eqs.(\ref{eqmo2}) implies that for $R$ small
$\dot{\phi}$ has a sign independent of $\phi$, namely the same as
the sign of the coefficient $d_0=a_{10}b_{01} -a_{01}b_{10}$.
Generically we have $d_0\neq 0$. This implies that $\phi$
describes rotations clockwise, if $d_0<0$, or anticlockwise, if
$d_0>0$. Furthermore, the angular frequency near the nodal point
scales as $\dot{\phi}=O(R^{-2})$.

The flow lines of Eqs(\ref{eqmo2}) close to the nodal point (for
$R$ small) are given by
\begin{equation}\label{drdf}
{dR\over d\phi}={c_2R^2+c_3R^3+c_4R^4+...\over
d_0+d_1R+d_2R^2+...} = {c_2\over d_0}R^2 + \bigg({c_3\over
d_0}-{c_2d_1\over d_0^2}\bigg)R^3+...
\end{equation}
The coefficient $c_2$ contains only terms of third degree in the
trigonometric functions $\sin\phi$, $\cos\phi$. Thus, averaging
Eq.(\ref{drdf}) over periods of the angle $\phi$ (which is fast
for $R$ small, $\dot{\phi}=O(1/R^2)$) yields
\begin{equation}\label{drdfav}
{d\bar{R}\over d\phi}=<f_3>\bar{R}^3+...
\end{equation}
where the coefficient $<f_3>$ is given by
$$
<f_3>(a_{ij},b_{ij},V_x,V_y)={1\over 2\pi}
\int_{0}^{2\pi}\bigg({c_3\over d_0}-{c_2d_1\over d_0^2}\bigg)d\phi
$$
with $i+j=0,1,2$. As explicitly demonstrated in Appendix A, for
any {\it non-zero value} of the velocity of the frame of reference
$(V_x,V_y)\neq (0,0)$ we have $<f_3>\neq 0$. Then the solutions of
Eq.(\ref{drdfav}) define {\it spirals}, i.e.
\begin{equation}\label{spsol}
\bar{R}(\phi)={R_0\over \sqrt{1-2R_0^2<f_3>(\phi-\phi_0)}}~~.
\end{equation}
This is a spiral terminating at $R=0$, i.e., at the nodal point,
when $\phi\rightarrow\infty$ (if $<f_3>~<~0$), or $\phi\rightarrow
-\infty$ (if $<f_3>~>~0$). Thus, depending on the sign of $<f_3>$
the nodal point is either an attractor or a repellor. The only
exception is when $(V_x,V_y)=(0,0)$, i.e. the motions are
considered in the rest frame. In that case we have $<f_j>=0$ for
all $j\geq 3$, i.e. the nodal point is a center. This follows
trivially from the condition $\nabla\cdot j=0$ implying that if we
set $H=-\int j_v du$, the components of the current are given by
$j_u=\partial H/\partial v$, $j_v=-\partial H/\partial
u$, which is equivalent to a Hamiltonian system $(du/ds,dv/ds)
\equiv(j_u,j_v)$ under the non-uniform time parametrization
$ds=G^{-1}dt=|\psi^{-2}|dt$. Thus, in the rest frame the nodal
point cannot be the limit of a spiral but it is a center of the
instantaneous flow (see also \cite{berry2005}).

\subsubsection{X-point} The X-point ($u_X,v_X$) is a second
critical point of the instantaneous flow found by setting
$du/dt=dv/dt=0$ at $(u,v)=(u_X,v_X)$. We find:
\begin{equation}\label{dudv}
{V_x\over V_y}=\frac{A v_X + B_1 u_X^2 + C_1 v_X^2 + D_1
u_X~v_X+\ldots} {-A u_X+ B_2 u_X^2 + C_2 v_X^2 + D_2
u_X~v_X+\ldots}
\end{equation}
where the coefficients $A,B_i,C_i$ and $D_i$ are readily derived from
Eqs.(\ref{eqmoexp}). If $(u_X,v_X)$ are small, we find an
approximate expression:
\begin{equation}\label{vxvy}
{V_x\over V_y}\simeq-\frac{v_X}{u_X}
\end{equation}
which, upon substitution to the first of Eqs.(\ref{eqmoexp})
yields a second order equation for, say, $u_X$. The non-zero
solution reads:
\begin{equation}\label{1v3}
u_X\simeq\frac{g_1
V_xV_y}{g_2V_x^2+g_3V_y^2+g_4V_xV_y+g_5V_x^3+g_6V_x^2V_y+g_7V_xV_y^2},
~~~v_X\simeq-{V_x\over V_y}u_X
\end{equation}
where all the coefficients $g_i$ depend only on the coefficients
$a_{ij},b_{ij}$. In particular:
$$
g_1=2(a_{10}b_{01}-a_{01}b_{10}),~~~
g_2=2a_{11}b_{01}+a_{10}b_{02}-a_{02}b_{10}-2a_{01}b_{11},~~~
g_3=a_{10}b_{02}-a_{02}b_{10},
$$
$$
g_4=2a_{02}b_{01}-2a_{01}b_{02},~~~ g_5=2(a_{01}^2+b_{01}^2),~~~
g_6=-4(a_{01}a_{10}+b_{01}b_{10}),~~~ g_7=2(a_{10}^2+b_{10}^2)~~.
$$
In numerical applications, Eq.(\ref{1v3}) is used to find a good
initial guess for the position of the X-point, while better
approximations are found by successive iterations of a
root-finding algorithm (e.g. Newton-Raphson) for the roots of the
system of equations (\ref{eqmouv}).

The linearized equations of motion around $(u_X,v_X)$ correspond
to the linear system formed by the Jacobian matrix $J[(\partial
S/\partial u,\partial S/\partial v),(u,v)]$ which is a $2\times 2$
symmetric matrix with constant coefficients. Thus the eigenvalues
$\lambda_1,\lambda_2$ are real. In the limit of small $u_X,v_X$,
the eigenvalues have opposite sign, since one readily finds that
$\lambda_1\lambda_2=-A^2/G_2^2 + O(1/R^3)$, where $G_2=O(R^2)$ is
the quadratic part of $G$ (Eq.(\ref{gi})). Hence $(u_X,v_X)$ is an
X-point with one unstable and one stable direction. Finally, the
measure of both eigenvalues scales as an inverse power of the
distance $R_X=\sqrt{u_X^2+v_X^2}$. Numerically we find (see EKC)
that the power-law scaling is $\lambda\sim R_X^{-p}$ with
$p\simeq 1.5$.

The numerator of the first of Eqs.(\ref{1v3}) is a $O(V^2)$
quantity, $V=\sqrt{V_x^2+V_y^2}$, while the denominator contains
both $O(V^2)$ and $O(V^3)$ quantities. If $V$ is small we have
$R_X=O(1)$. This, as shown below, sets an upper limit of validity
of the adiabatic approximation for $R_X$. On the other hand, if
$V$ is large we have $R_X=O(V^{-1})$.

\subsection{Conditions of validity of the adiabatic approximation}

Conditions for the validity of the adiabatic approximation are now
visualized with the help of Figure 1 (schematic), showing the nodal
point - X-point complex as viewed in a frame of reference of arbitrary
velocity $(V_x(t),V_y(t))$ at two nearby times $t=t_0$ (solid lines)
and $t'=t_0+\Delta t$ (dashed lines). $R_X$ is the distance of the
X-point from the nodal point, while $\Delta R_0$ is the distance
traveled by the nodal point with respect to this particular frame
of reference within the time interval $\Delta t$. The vector
$\vec{\mbox{v}}$ refers to the velocities of the particles' orbits
as seen in the moving frame of reference.
In the adiabatic approximation the velocities $\vec{\mbox{v}}$ must
be large enough so that the flow integral curves change slowly relatively
to the change of a particle's position within the time interval
$\Delta t$. Since $G$ in Eqs.(\ref{eqmoexp}) depends, to the lowest order,
quadratically on $u,v$ (Eq.(\ref{gi})), the particles' velocities become
arbitrarily large if $u,v$ become sufficiently small. The linear
velocities at a distance $R$ from $(x_0,y_0)$ scale as $|v|\sim
R^{-1}$. Very close to $(x_0,y_0)$ the motions are spiral-like
with a frequency of order $\omega\sim |v|/R\sim R^{-2}$, or period
$T\sim R^2$. The linear size of the `nodal point - X-point
complex' can be estimated by the distance $R_X$ which is of order
$R_X\sim V^{-1}$ (Eq.(\ref{1v3})). The shift of the nodal point
$\Delta R_0$ within one period is estimated as $\Delta R_0\sim
T|\vec{V}-\vec{V}_0|$, where $\vec{V}_0$ is the velocity of the
nodal point in the rest frame. For the adiabatic approximation to
hold, the following two conditions are sufficient and necessary:
\begin{figure}\label{bifurc}
\centering
\includegraphics[scale=0.5]{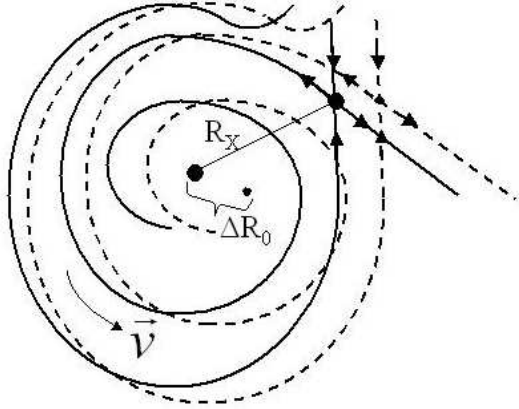}
\caption{\small Schematic representation of the `nodal point -
X-point' complex as viewed in a frame of reference of arbitrary
velocity $(V_x(t),V_y(t))$ at two nearby times $t=t_0$ (solid
lines) and $t'=t_0+\Delta t$ (dashed lines). $\Delta R_0$ is the
length traveled by the nodal point within the time step $\Delta t$.
$R_X$ is the distance from the nodal point to the X-point at
$t=t_0$.}
\end{figure}


a) The shift $\Delta R_0$ must be small with respect to the linear
size $R_X$ of the nodal point - X-point complex. This condition
yields $\Delta R_0<<R_X$, or $R_X^2|\vec{V}-\vec{V}_0|<<R_X$
implying
\begin{equation}\label{adcon1}
{|\vec{V}-\vec{V}_0|\over V}<<1~~.
\end{equation}
Thus the first condition is that $\vec{V}\simeq\vec{V}_0$, i.e.
the frame velocity $\vec{V}$ should be close to the velocity
$\vec{V}_0$ of the nodal point with respect to the rest frame.

b) The characteristic velocities within the `nodal point - X-point
complex' (i.e. for $R<R_X$) must be much larger than the rate of change
of the coefficients $a_{ij},b_{ij}$. Generically, the rates of change
of $a_{ij},b_{ij}$ are in general $O(1)$ quantities (depending
on trigonometric functions of the time and on the wavefunction's
normalized amplitudes, see for example EKC). Thus $|\vec{v}|$
should satisfy $|\vec{v}|>1$, or, since $|\vec{v}|\sim
R^{-1}>R_X^{-1}$,
\begin{equation}\label{adcon2}
R_X\sim V^{-1}< 1~~.
\end{equation}
Thus, the second condition is that the velocity of the moving frame
of reference should be large with respect to the rest frame. In particular,
the rest frame itself represents a frame in which the use of the adiabatic
approximation is not, in general, valid.

\subsection{Local Lyapunov exponents of scattered orbits}

The close encounters of the orbits with the nodal point - X-point
complex can be approximated as {\it scattering events}, in which
an orbit approaches the complex in a direction close to the stable
manifold of the X-point and recedes from the complex in a direction
close to the unstable manifold of the X-point. A simple model
to describe the scattering is obtained by noting that if
(without loss of generality) the axes are rotated so that
at $t=t_0$ the velocity of the nodal point is along the x-axis,
i.e. $\dot{x}_0\neq 0$, $\dot{y}_0=0$, an expansion of the form
(\ref{psiexp}) in the new coordinates yields the equations of motion
\begin{equation}\label{eqrot}
{du\over dt}={-Av+\ldots\over Bu^2+2Cuv+Dv^2+\ldots}-\dot{x}_0,~~~
{dv\over dt}={Au+\ldots\over Bu^2+2Cuv+Dv^2+\ldots}
\end{equation}
where $A=a_{10}b_{01}-a_{01}b_{10}$, $B=a_{10}^2+b_{10}^2$,
$C=a_{01}a_{10}+b_{01}b_{10}$, $D=a_{01}^2+b_{01}^2$. The
quadratic form in the denominator is always positive definite.
The X-point is on the $v-$axis, with $(u_x,v_x)=(0,-AD^{-1}
\dot{x}_0^{-1})$, consistent with the scaling $R_x\sim V^{-1}
=|\dot{x}_0|^{-1}$ found in the previous subsection. Ignoring the
higher order terms in the numerators of Eqs.(\ref{eqrot}) causes the
nodal point to become a center rather than the limit of a
spiral. This poses however no problem to the study of orbits
being scattered by the nodal point - X-point complex since these
orbits avoid penetrating the interior of the separatrix
domain, close to the nodal point (see next subsection).
Eq.(\ref{eqrot}) is then suggestive of the following simple model
\begin{equation}\label{eqrot2}
{du\over dt}={-v\over u^2+v^2}-\dot{x}_0,~~~ {dv\over dt}={u\over
u^2+v^2}
\end{equation}
to describe the scattering of orbits by the nodal point - X-point
complex.

The flow (\ref{eqrot2}) admits the integral
\begin{equation}\label{ccon}
C=e^{2\dot{x}_0v}(u^2+v^2)
\end{equation}
which is obeyed by the scattered orbits locally, as long as the
scattering lasts. The time evolution of $R=\sqrt{u^2+v^2}$ is given
by $R dR/dt=\sqrt{\dot{x}_0^2R^2-(\ln R-{1\over 2}\ln
C)^2}$. Far from the complex Eqs.(\ref{eqrot2}) take the form
$du/dt\approx -\dot{x}_0$, $dv/dt\simeq 0$. Thus $d(R^2)/dt\approx
-2\dot{x}_0(u_0-\dot{x}_0 t)$, where $u_0=R_0$ is the initial
condition of an orbit on the u-axis ($v_0=0$), implying
\begin{equation}\label{r2rot}
R^2\approx R_0^2-2\dot{x}_0u_0t +\dot{x}_0^2t^2~~.
\end{equation}
The orbits passing outside the separatrix loop of the nodal point -
X-point complex can be divided into 'type I' and 'type II' orbits
(Figure 2a). Type I orbits surround the separatrix loop, while
Type II orbits pass below the X-point, not surrounding the separatrix
loop. In either case, the average time of a scattering event can be
estimated by the non-trivial root for $t$ of Eq.(\ref{r2rot}) with
$R=R_0$, namely
\begin{equation}\label{tsc}
t_{scatter} = O\bigg({2u_0\over \dot{x}_0}\bigg)~~.
\end{equation}
\begin{figure}\label{toysep}
\centering
\includegraphics[scale=0.8]{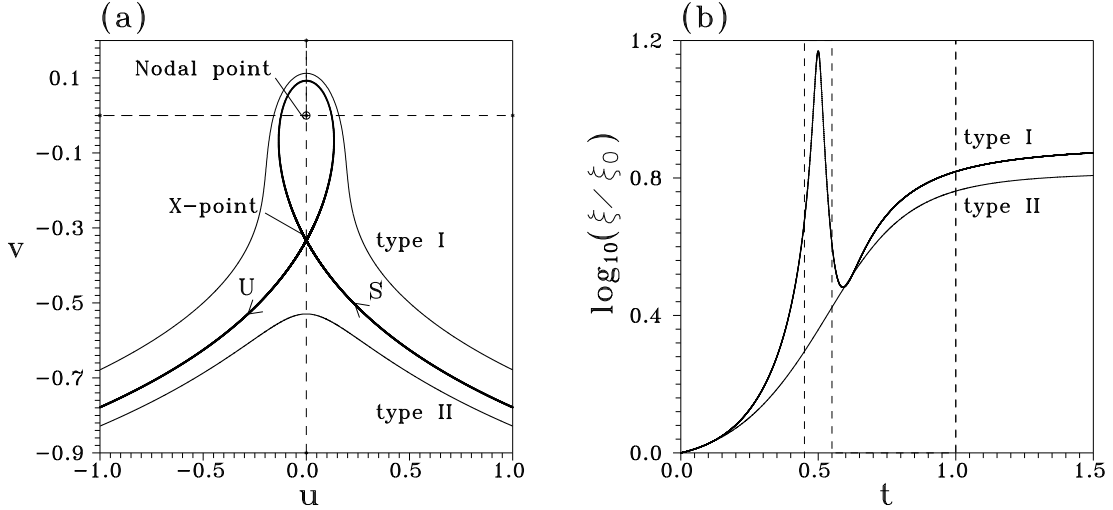}
\caption{\small (a) The thick looped curve is the separatrix
passing from the X-point of the model (\ref{eqrot2}) when
$\dot{x}_0=3$. The stable and unstable manifolds are marked S and
U respectively. The stable manifold crosses the line $u=1$ at
$v_s=-0.7785019...$. The upper and lower thin solid lines
represent a `type I' orbit (initial conditions $u=1, v=v_s+0.01$)
and a `type II' orbit (initial conditions $u=1, v=v_s-0.005$)
respectively. (b) The time growth of the deviations $\xi(t)$ for
the type I (thick curve) and type II (thin curve) orbits. In both
cases the initial deviation vector is taken $\vec{\xi}_0=(1,0)$.
The rightmost vertical dashed line at $t=1$ corresponds
approximately to the time when the orbits reach $u=-1$, i.e. a
position symmetric to their initial conditions with respect to the
$v=0$ axis. The two left vertical dashed lines mark the time
window $0.45\leq t\leq 0.55$ within which the type I orbit forms
part of a loop around the nodal point. }
\end{figure}

Deviations of $t_{scatter}$ from the estimate of Eq.(\ref{tsc})
take place when the orbits are very close to the invariant
manifolds of the X-point, since $t_{scatter}\rightarrow\infty$ as
the initial conditions tend to a point on the stable manifold S.
As shown in the Appendix B, such deviations lead to increased local
Lyapunov characteristic numbers of the scattered orbits. The time evolution
of the length of the deviation vector $\vec{\xi}(t)=(\Delta u,
\Delta v)$ has a characteristic `profile' along the scattering,
different for type I or type II orbits (Figure 2b). In the case of
type I orbits, $\xi(t)$ exhibits a rise and fall phase during the
description of the separatrix loop. This phase lasts for a time
$t_{loop}$ estimated as $t_{loop}\sim R_x^2\sim |\dot{x}_0|^{-2}$.
If $|\dot{x}_0|>>1$ we have $t_{loop}<<t_{scatter}$. Most of the
growth of $\xi$ takes place after $t=t_{loop}$, as the orbit recedes
along the X-point's unstable manifold. In the case of type II
orbits the $\xi(t)$ time profile exhibits a continuous rise
from the start and $\xi(t)$ tends to stabilize after
$t=t_{scatter}$.

\begin{figure}\label{toysep}
\centering
\includegraphics[scale=0.8]{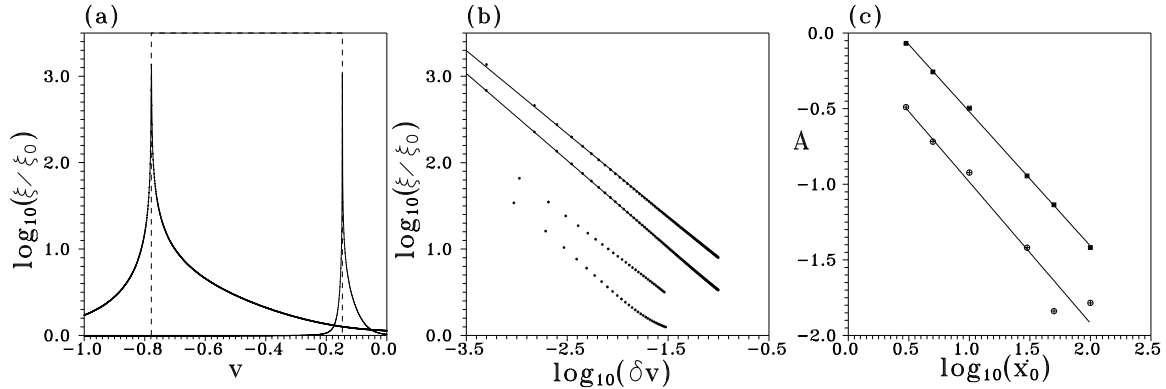}
\caption{\small (a) The final value of $\xi/\xi_0$ (on a
logarithmic scale) as a function of the initial condition $v=v_1$ of
the orbits of the model (\ref{eqrot2}) taken on the line $u=1$ at $t=0$.
The left and right curves correspond to $\dot{x}_0=3$ and
$\dot{x}_0=30$ respectively. The vertical dashed lines mark the position
$v=v_s$ at which the X-point's stable manifold S crosses the line $u=1$
in each case. (b) $\xi/\xi_0$ as a function
of $\delta v_1 = |v_1-v_s|$ on a logarithmic scale. The two top
curves correspond to $\dot{x}_0=3$ (upper curve for type II
orbits, lower curve for type I orbits). The straight solid lines
passing through these curves represent a power-law fitting
$\xi/\xi_0 = A\delta v_1^{-b}$ with $b=0.95..$ for the upper curve
and $b=1.01$ for the lower curve. The bottom two curves correspond
to $\dot{x}_0=30$. (c) A power-law fitting of the constant $A$ as
a function of $\dot{x}_0$ ($A=0.89 \dot{x}_0^{-0.93}$  and
$A=2.33\dot{x}_0^{-0.89}$ for the lower (type I) and upper (type
II) curves respectively). }
\end{figure}
A theoretical quantitative estimate of the local value of the
Lyapunov characteristic number in a scattering event is made in Appendix
B. The growth of deviations is modeled by calculating the differential
velocity of motions in two nearby integral curves of the flow
(\ref{eqrot2}) close to the X-point's stable and unstable manifolds.
This modeling yields the scaling law
\begin{equation}\label{xiovx0}
{\xi\over\xi_0}\sim {1\over \dot{x}_0\delta v_1}
\end{equation}
where $\xi_0$ and $\xi$ are the lengths of the deviation vectors
of a scattered orbit before and after the scattering respectively,
and $\delta v_1$ is the initial distance of the orbit from the
X-point's stable manifold far from the complex. The latter quantity
is called the `impact parameter'. The theoretical prediction
(\ref{xiovx0}) is well reproduced numerically, by taking many
trajectories in the model (\ref{eqrot2}), for different values
of $\dot{x}_0$ (Figure 3).

Since the local eigenvalue $\lambda$ of the linearized flow near
the X-point scales as a positive power of $\dot{x}_0$, Eq.(\ref{xiovx0})
implies that the chaotic scattering takes place mainly in
encounters in which $\lambda$ is relatively small (though non zero).
This appears at first
counter-intuitive. However, even if a trajectory is started close
to the asymptotic manifolds of the X-point, it is in general far
from the X-point itself, except for a short transit time of order
$\sim 1/\dot{x}_0^2$. Thus, to describe the total scattering
correctly one has to take into account nonlinear terms of the expansion
of the equations of motion, which introduce large deviations from the
locally hyperbolic dynamics. On the other hand, the whole previous
analysis relies on the use of the adiabatic
approximation, which, according to subsection II B, holds better
when $\dot{x}_0$ is large. For a nodal point - X-point complex to
cause effective chaotic scattering, we thus have both an upper
and lower restriction to the values of $\dot{x}_0$. Precise
upper and lower limits on $\dot{x}_0$, or, equivalently, the size
of a complex $R_X\sim 1/\dot{x}_0$, for the complexes to produce
effective chaotic scattering, can only be found by numerical
experiments, as substantiated by specific examples in subsection
III D.

\section{Numerical study}

\subsection{A numerical example of the nodal point - X-point complex}

In our previous paper (EKC) we studied the `nodal point - X-point'
complex in a system of two harmonic oscillators
\begin{equation}\label{ham2dharm}
H={1\over 2}(p_x^2+p_y^2) + {1\over 2}(x^2 + (c y)^2)
\end{equation}
when the guiding field is the superposition of the ground state
and the two first excited states \cite{pv1995}
\begin{equation}\label{eigenharm}
\psi(x,y,t) = e^{-{x^2+cy^2\over 2}-i{(1+c)t\over 2}}\big(
1+axe^{-it}+bc^{1/2}xye^{-i(1+c)t}\big)~~
\end{equation}
while the frequencies are incommensurate, $\omega_1=1$,
$\omega_2=c=\sqrt{2}/2$. If we select a moving frame of reference such
that its center $(x_0(t),y_0(t))$ coincides at all times with the moving
nodal point, Eqs.(\ref{eqmoexp}) take the form:
\begin{eqnarray}\label{equv1}
{du\over dt} &= &-{bc^{1/2}v\sin(1+c)t\over G}-\dot{x}_0 \\
{dv\over dt} &= &{bc^{1/2}u\sin(1+c)t-abc^{1/2}u^2\sin ct\over
G}-\dot{y}_0  \nonumber
\end{eqnarray}
where $G=G_2+G_3+G_4$ with $G_2 = (u^2/x_0^2)-2bc^{1/2}uv\cos(1+c)t+b^2cx_0^2v^2$,
$G_3=-(2bc^{1/2}/x_0)u^2v\cos(1+c)t+2b^2cx_0uv^2$,
$G_4 = b^2cu^2v^2$ and $\dot{x}_0,\dot{y}_0$ are found by differentiating
$x_0(t)$,$y_0(t)$, which are given by
\begin{equation}\label{nodal}
x_0(t) = -{\sin(1+c)t\over a\sin ct},~~~~ y_0(t) = -{a\sin t\over
bc^{1/2}\sin (1+c)t}~~.
\end{equation}

Figure 4 shows examples of the form of the nodal point - X-point
complex in the above system, in a specific time interval. In all
cases four asymptotic manifolds start from the X-point along pairs
of opposite, stable or unstable, directions. One asymptotic
manifold goes towards the nodal point in a spiral way and the
other three extend to infinity. The nodal point itself acts as an
attractor or a repellor, and an asymptotic curve starting at the
nodal point forms a spiral outwards. The sense of motion around
the nodal point is determined by the sign of $<f_3>$
(Eq.(\ref{drdfav})), which is given in this case by \footnote{ In
our previous paper (EKC) an error appears in the second factor of
$<f_3>$ in Eq.(35). This equation is derived from Eq.(A4) which is
correct. However, in Eq.(35) the term $x_0\dot{x}_0$ should be
replaced by ${x_0\dot{x}_0(1-b^2cx_0^4)\over (1+b^2cx_0^4)}$. The
numerical results change only slightly. Note that Fig.11b of EKC
gives only the second factor of $<f_3>$, that changes sign. Three
more typos have been found in the caption of Fig.12, namely the
integer part of $t_0$ in the cases (a),(b),(c) is 175, as in case
(d).}
\begin{eqnarray}\label{f3}
<f_3> &= &\Bigg({1+b^2cx_0^4\over 4bc^{1/2}x_0^4\sin(1+c)t_0}\Bigg)\times\nonumber\\
& &\Bigg( {1-b^2cx_0^4\over 1+b^2cx_0^4} x_0\dot{x}_0
+{\dot{x}_0\dot{y}_0(b^2cx_0^4-1)\over bc^{1/2}\sin(1+c)t_0}
-x_0^2(\dot{x}_0^2-\dot{y}_0^2)\cot(1+c)t_0 \Bigg)~.
\end{eqnarray}

In view of Eq.(\ref{spsol}), if $<f_3><0$ the value of $R$
decreases towards $R=0$ with increasing $\phi$, therefore the
nodal point is an attractor if the spiral is described
counterclockwise and a repellor if it is described clockwise. The
opposite is true if $<f_3>~>0$. On the other hand, the coefficient
$d_0$ in Eq.(\ref{eqmo2}) turns out to be equal to
$d_0=\sin(1+c)t_0$ (while $G$ is positive for any
$(u,v)\neq(0,0)$. Thus, if $\sin(1+c)t_0>0$ (or$<0$) the spiral is
described counterclockwise (or clockwise). When $<f_3>=0$ (and
$\sin(1+c)t_0\neq 0$) the nodal point changes from an attractor to
a repellor. Then we have a Hopf bifurcation, followed by the
formation of a limit cycle.

\begin{figure}\label{toysep}
\centering
\includegraphics[scale=0.7]{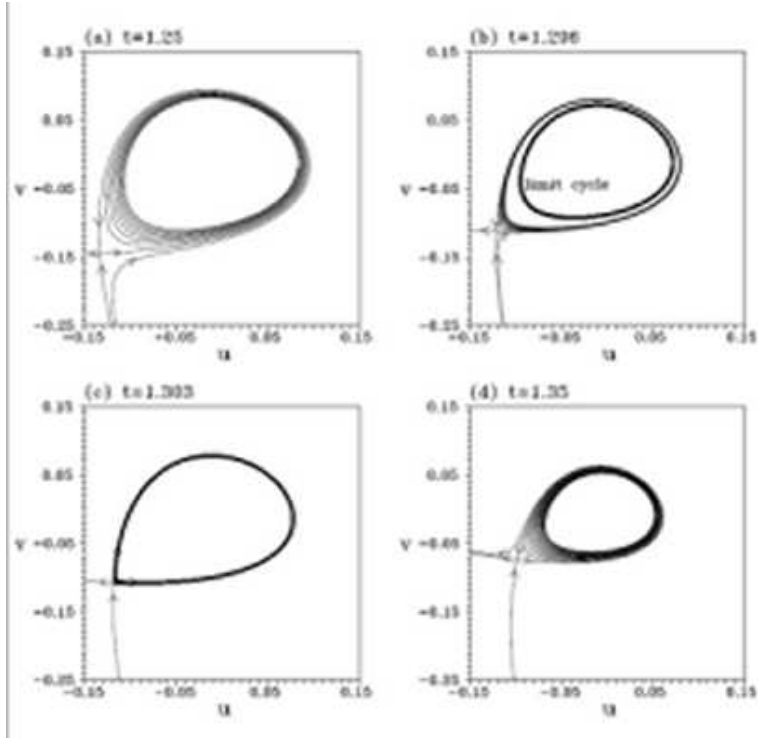}
\caption{\small The form of the nodal point - X-point complex in
the EKC model (equations of motion given by (\ref{equv1}) with $a=b=1$,
$c=\sqrt{2}/2$) for
the times indicated in panels (a) to (d). The nodal point is an
attractor at $t=1.25$ (a). A Hopf bifurcation takes place near
$t=1.29415$. The nodal point becomes a repellor  and the associated
limit cycle moves outwards at subsequent times. E.g. at $t=1.296$
it has the position shown in (b). At $t=1.303$ the limit cycle
reaches the X-point (c). Then the limit cycle disappears and all
the orbits inside the nodal point - X-point complex are repelled
away from the complex (e.g. at $t=1.35$,(d))}.
\end{figure}
\begin{figure}\label{f3val}
\centering
\includegraphics[scale=0.9]{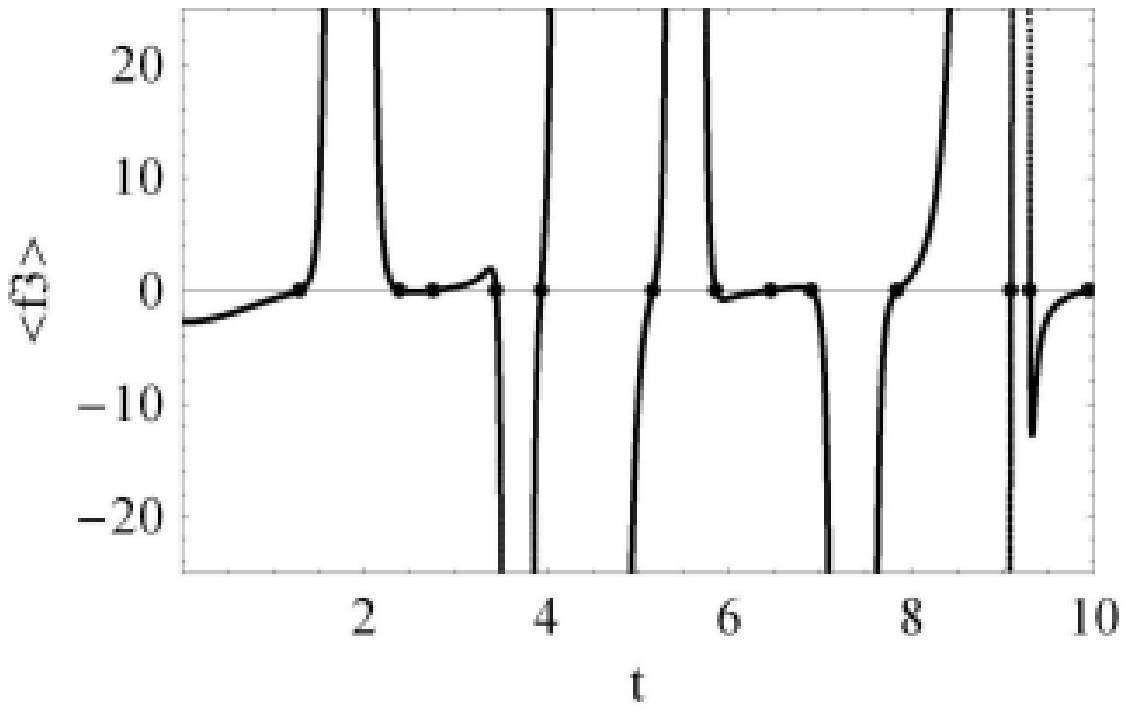}
\caption{\small The value of $<f_3>$ as a function of the time $t$
in the interval $0\leq t\leq 10$ in the EKC model with $a=b=1$,
$c=\sqrt{2}/2$. The dots mark the points where $<f_3>=0$. }
\end{figure}
As an example we consider the evolution of the manifolds between
$t=1.25$ and $t=1.35$ (Fig.4). For $t=1.25$ (Fig.4a) we have
$<f_3>~<0$ and $\sin(1+c)t>0$ therefore the nodal
point is an attractor and the orbits appoaching it are spirals
described counterclockwise. One orbit of this type is one of the
unstable manifolds of the X-point. The other unstable manifold and
the two stable manifolds of the X-point extend to infinity. In
particular, the upper stable manifold escapes downwards
after making an almost complete rotation (backwards in time)
clockwise around the nodal point.

The attraction of the orbits by the nodal point terminates
when a transition of $<f_3>$ takes place from negative to positive,
near $t=1.294$. Then, the nodal point becomes a repellor and a
limit cycle is formed around it. The limit cycle moves outwards
(e.g. $t=1.296$, Fig.4b). The orbits both inside the limit cycle
(i.e. close to the nodal point) and outside the limit cycle
(between the limit cycle and the X-point) are attracted by the
limit cycle. The orbits can enter the complex only via a very
narrow channel formed by the two stable manifolds below the
X-point. As the limit cycle moves outwards and approaches
the X-point, any orbit approaching the limit cycle is dragged
closer and closer to the X-point.

The limit cycle reaches the X-point at $t=1.303$ (Fig.4c).
Then the unstable manifold from the right joins the upper
stable manifold  and together they form a separatrix.
For still larger $t$ (e.g. $t=1.35$, Fig.4d) the limit cycle
has disappeared and the upper stable manifold approaches the
nodal point via spiral rotations (backwards in time).
On the other hand, the two unstable manifolds go to infinity
on the left, one directly, and the other after an almost complete
rotation (counterclockwise) around the nodal point.

The transition displayed in Fig.4 constitutes a Hopf bifurcation.
Hopf bifurcations take place whenever $<f_3>$ crosses a zero value.
The bifurcation displayed in Fig.4 is called direct (the limit
cycle is formed first near the nodal point and later it disappears
at a separatrix). However, inverse Hopf bifurcations are also
commonly observed, in which the limit cycle moves towards the
nodal point. The rate of appearance of direct or inverse
Hopf bifurcations is a few per period (which is of order
$2\pi$). A typical survival time for limit cycles is
$\Delta T\approx 10^{-2}$ (in Fig.4 we have $\Delta T=0.008$).

The value of $<f_3>$ follows a time evolution as exemplified in
Fig.5. The value of $<f_3>$ becomes infinite when $\sin(1+c)t_0=0$
and when $\sin ct_0=0$, with $t_0\neq 0$. In Fig.5 $<f_3>=±\infty$
at times $t_0=k\pi/(1+c)$ with $k=1\ldots 5$ and $t_0=\pi/c$ and $2\pi/c$.
Between $k=1$ and $2$ the sign of $\sin(1+c)t_0$ is negative, thus
the nodal point is a repellor whenever $<f_3>~>0$. Between $k=2$ and $3$
the sign of $\sin(1+c)t_0$ is positive, thus the nodal point is a
repellor whenever $<f_3>~<0$, and so on. Between two points with
$<f_3>=±\infty$ there may be a number of times where $<f_3>=0$
(one, two, or three times in Fig.5). Evolutions of the phase
portraits similar to Fig.4 appear close to all the time moments
when $<f_3>=0$.

Using the above rules, in the time interval $0\leq t\leq 10$ the
nodal point is found to be a repellor for about  $47\%$ of the total
time. Whenever the nodal point is a repellor it cannot in general be
approached by any orbits. A rare exception is the case in which an
inverse Hopf bifurcation takes place at the nodal point. Then the
orbits which are initially in an extremely narrow channel of the flow,
formed between the two {\it stable} manifolds outside the X-point,
approach the limit cycle which tends to the nodal point. Such events
can only last for times $\Delta T\approx 10^{-2}$.

Similarly, when the nodal point becomes an attractor, only an extremely
narrow channel of the flow formed by the stable manifolds  allows
for the orbits to go deeply inside the complex, i.e., to approach
the nodal point. But this channel also disappears in transient
time intervals $\Delta T\approx 10^{-2}$ in which the nodal
point is protected by a limit cycle.

We conclude that the asymptotic curves of the X-point in most
cases {\it do not allow} close approaches to the nodal point.
Only if two conditions are satisfied, namely that (a) the inner
asymptotic curve is unstable, and (b) the nodal point (or a
limit cycle approaching it) is an attractor, we may have close
approaches to the nodal point. In all cases, however, the approach
is only possible for a set of initial conditions of extremely
small measure, i.e., the orbits in general avoid the nodal
point.

\subsection{Type I and Type II interactions}

Figures 6 and 7 show now examples of Type I (Fig.6) and Type II
(Fig.7) interactions of a quantum trajectory with the nodal point
- X-point complex in the above system. The trajectories are viewed
in the moving frame of reference $(u,v)$ centered at the nodal point
and they are superposed to the background instantaneous velocity
flow at the indicated times $t$. The deviation vector $\vec{\xi}$
is calculated numerically by the variational equations of motion.
Its local direction is indicated by the thick arrows in each
panel, while the time evolution of the normalized length $\xi/\xi_0$
is shown in the last panels of Figs.6 and 7. These two curves
are compared to the theoretical curves of Fig.2b for type I and
type II events respectively. Note in particular that in the case
of the type I event the growth of $\xi$ during the description of
the first half-loop is nearly compensated by a decrease in the
second half-loop, and the deviations start growing essentially
after the description of the loop. The two local peaks of $\xi(t)$
in the interval of the loop description can be explained by the
fact that the loop is not perfectly circular (see \cite{cve1997}
for an explanation of the behavior of $\xi$ in
non-circular invariant curves). Also, in the case of the type
II orbit the growth of $\xi$ does not take place very close to
to the X-point, but all along the scattering event, in accordance
with the theory of section II.
\begin{figure}\label{scattera}
\centering
\includegraphics[scale=1.3]{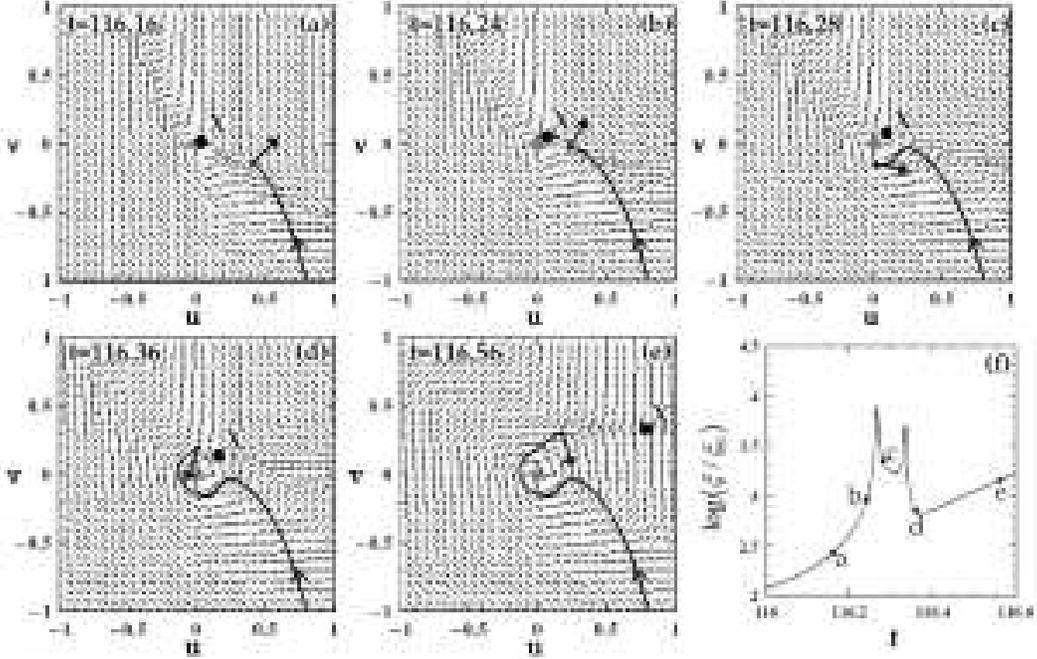}
\caption{\small A type I event in the EKC model with $a=b=1$,
$c=\sqrt{2}/2$, and initial conditions of the trajectory
$x_0=1.71510$, $y_0=1.29285$ at $t=0$. In panels (a) to (e), the
nodal point appears as a gray thick dot at $(0,0)$, while the
X-point appears as a black thick dot (denoted X). The background
small arrows indicate the local instantaneous velocity flow in the
$(u,v)$ frame of reference. The thick long arrow shows the {\it
direction} of the deviation vector $\vec{\xi}$ for the same orbit
(initial conditions $\vec{\xi}_0=(1,0)$). Panel (f) shows the time
evolution of the normalized length $\xi/\xi_0$ on a logarithmic
scale.}
\end{figure}
\begin{figure}\label{scatterb}
\centering
\includegraphics[scale=1.3]{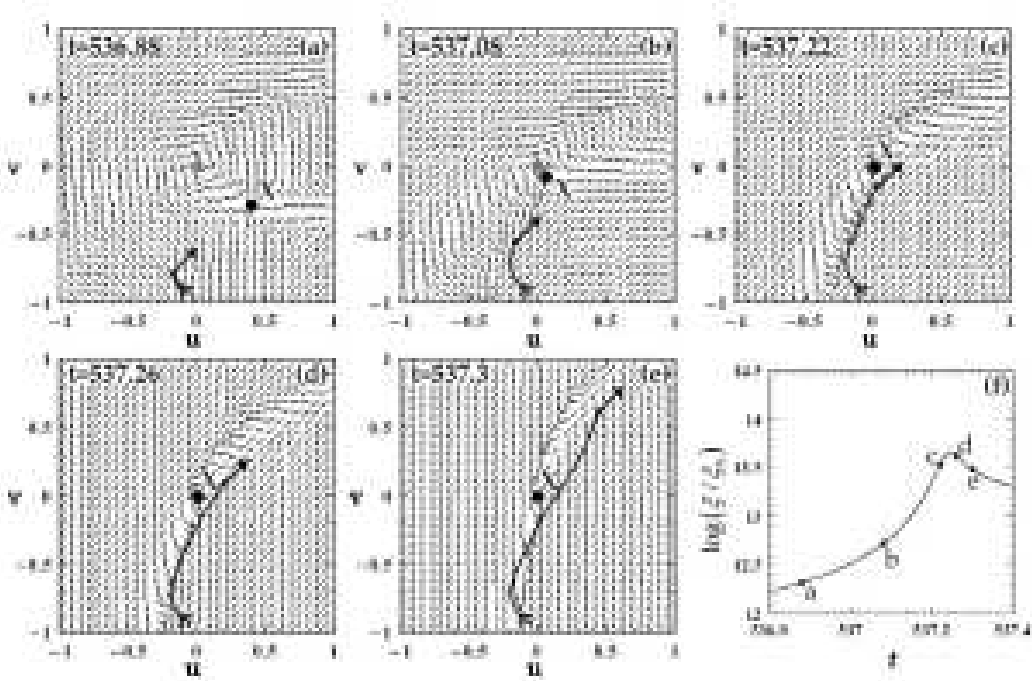}
\caption{Same as in Fig.6 for a type II event of the same orbit.}
\end{figure}

\subsection{Multiple nodal points}

In EKC we considered the orbits in simple $\psi-$fields of the
system of two harmonic oscillators given by the Hamiltonian
(\ref{ham2dharm}). The eigenfuctions are
\begin{equation}\label{eqn:02}
\Psi_{n_1 n_2}=e^{-iE_{n_1 n_2}t}e^{-x^2/2}H_{n_1}\left(x\right)
e^{-cy^2/2}H_{n_2}\left(\sqrt{c} y\right)
\end{equation}
where the energy of the state $(n_1n_2)$ is
\begin{equation}\label{eqn:03}
E_{n_1n_2}=\left(\frac{1}{2} +n_1\right)\omega_1
+\left(\frac{1}{2} +n_2\right)\omega_2
\end{equation}
and $H_n$ are Hermite polynomials. The following eigenfunctions
are explicitly referred to in the sequel:
\begin{eqnarray}\label{eigpsi}
\Psi_{00}&=&e^{-\frac{i}{2}\left(1+c\right)t}e^{-\frac{\left(x^2+cy^2\right)}{2}},~
\Psi_{10}=e^{-\frac{i}{2}\left(3+c\right)t}e^{-\frac{\left(x^2+cy^2\right)}{2}}x,~
\Psi_{11}=e^{-\frac{3i}{2}\left(1+c\right)t}e^{-\frac{\left(x^2+cy^2\right)}{2}}x\sqrt{c}y\nonumber\\
\Psi_{20}&=&e^{-\frac{i}{2}\left(5+c\right)t}e^{-\frac{\left(x^2+cy^2\right)}{2}}\left(x^2-1\right),~~~
\Psi_{30}=e^{-\frac{i}{2}\left(7+c\right)t}e^{-\frac{\left(x^2+cy^2\right)}{2}}\left(x^3-3x\right)~~.
\end{eqnarray}
In EKC we considered particular quantum trajectories in the case
$\Psi=\Psi_{00}+a\Psi_{01}+bc^{1/2}\Psi_{11}$
(Eq.(\ref{eigenharm})), with $a,b$ real, yielding one moving nodal
point. Here we shall consider the cases with a progressively higher
number of moving nodal points in the same Hamiltonian model. We call
`nodal lines' the trajectories of the nodal points. For the nodal
points to be moving, there are restrictions on the choice of
combination of the eigenfunctions, since for particular
combinations there are no nodal lines but isolated nodal points
appearing at specific times only. For example, if the wavefunction
consists of the sum of three eigenfunctions
$\Psi=\Psi_{n_1n_2}+a\Psi_{n_1^{'}n_2^{'}}+b\Psi_{n_1^{''}n_2^{''}}$
with equal quantum numbers $n_2=n_2^{'}=n_2^{''}$, the nodal
points $\Psi=0$ satisfy the equations
\begin{equation}\label{eqn:12}
\Psi=H_{n_1}\left(x\right)+ a
e^{-i\left(n_1^{'}-n_1\right)t}H_{n_1^{'}} \left(x\right)+ b
e^{-i\left(n_1^{''}-n_1\right)t}H_{n_1^{''}}\left(x\right)=0~~.
\end{equation}
Thus we have two distinct equations for the real and imaginary
parts of $\Psi$, implying that we have solutions for $x$ only at
specific times $t$. The same happens if $n_1=n_1^{'}=n_1^{''}$.
The same is true if we have more terms in $\Psi$, but with the
same quantum number $n_1$, or $n_2$, in all the terms. Such cases
are not examined below.

The nodal lines may enclose domains in the configuration space devoid
of nodal points for all times. In such domains the quantum
trajectories turn to be regular. Such empty domains are found
even if we take combinations of eigenfunctions yielding
simultaneously more than one nodal points. Examples are:
\begin{figure}\label{node00101120}
\centering
\includegraphics[scale=0.4,angle=0]{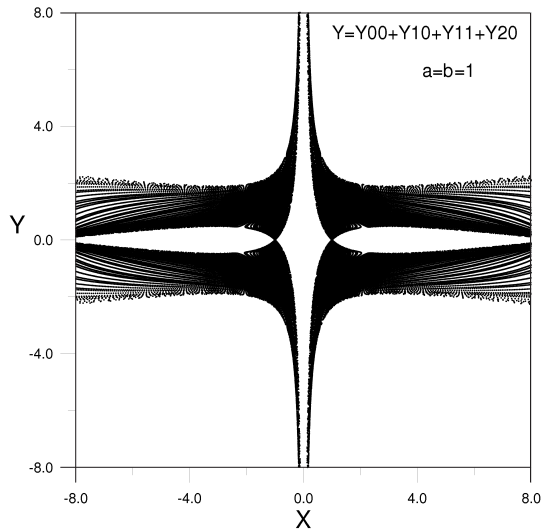}
\caption{\small Nodal lines of the wavefunction
$\Psi=\Psi_{00}+a\Psi_{10}+b\Psi_{11}+\epsilon\Psi_{20}$ when
$a=b=1$, $c=\sqrt{2}/2$, $\epsilon=0.1$.}
\end{figure}

\subsubsection{Case $\Psi=\Psi_{00}+a\Psi_{10}+b\Psi_{11}+\epsilon\Psi_{20}$}

If we add a fourth term of the form $\Psi_{20}$ in
Eq.(\ref{eigenharm}), i.e.:
\begin{equation}\label{eqn:31}
\Psi=\Psi_{00}+a\Psi_{10}+b\Psi_{11}+\epsilon\Psi_{20}~~
\end{equation}
the real and imaginary parts of the equation $\Psi=0$ take the
form
\begin{eqnarray}\label{eqn:33}
&~&1+ax\cos t+bx\sqrt{c}y\cos
\left(1+c\right)t+\epsilon\left(x^2-1\right)\cos 2t=0\nonumber\\
&~&ax\sin t +bx\sqrt{c}y\sin
\left(1+c\right)t+\epsilon\left(x^2-1\right)\sin 2t=0~~.
\end{eqnarray}
Multiplying the first equation  by $\sin(1+c)t$ and the second
equation by $\cos(1+c)t$ and subtracting, we find the equation
\begin{equation}
\sin \left(1+c\right)t+ax\sin ct+\epsilon\left(x^2-1\right)\sin
\left(c-1\right)t=0
\end{equation}
with solution
\begin{equation}\label{eqn:36}
x=\frac{1}{2\epsilon\sin \left(c-1\right)t}\{-a\sin ct \pm
\left[a^2\sin^2 ct-4\epsilon\sin \left(c-1\right)t \left(\sin
\left(1+c\right)t-\epsilon\sin
\left(c-1\right)t\right)\right]^{1/2}\}
\end{equation}
If $\sin ct$ is not close to zero and $\epsilon$ is small we find
\begin{eqnarray}\label{eqn:37}
x&=&\frac{1}{2\epsilon\sin \left(c-1\right)t}\{-a\sin ct \pm a\sin
ct
\bigg[1-\frac{2\epsilon\sin(c-1)t \sin(1+c)t}{a^2\sin^2 ct}\nonumber\\
&+&\frac{2\epsilon^2\sin^2 (c-1)t}{a^2\sin^2 ct}
-\frac{2\epsilon^2\sin^2 (c-1)t \sin^3(1+c)t}{a^4\sin^4
ct}\bigg]\}
\end{eqnarray}
If $x_0$ denotes the solution (\ref{nodal}), for $\epsilon=0$, the
solution close to $x=x_0$ for $\epsilon$ small is the one with the
plus sign
\begin{equation}\label{eqn:38}
x=-\frac{\sin(1+c)t}{a\sin ct}+\frac{\epsilon\sin (c-1)t}{a\sin
ct} \left[1-\frac{\sin^2(1+c)t}{a^2\sin^2 ct}\right]
=x_0+\frac{\epsilon\sin (c-1)t}{a\sin ct}(1-x_0^2)~,
\end{equation}
and
\begin{equation}\label{eqn:40}
y=\frac{1}{xb\sqrt{c}\sin (1+c)t}\left[-ax\sin
t+\epsilon(1-x^2)\sin 2t \right]~~.
\end{equation}
When $t=k\pi$ we have $y=0$ and $x=\pm\frac{1}{a}+O(\epsilon^2)$.
In particular if $\alpha=1$, we have $x=\pm 1$ exactly. This is
seen in Fig.8. Besides this solution we have $y=0$ also if
$-ax+2\epsilon(1-x^2)\cos t=0$, hence
$$
x=-\frac{a}{4\epsilon\cos
t}\left[1\pm\sqrt{1+\frac{16\epsilon^2\cos^2 t}{a^2}}\right]
$$
and if we take terms up to $O(\epsilon)$ we find
\begin{equation}\label{eqn:44}
x=\frac{2\epsilon}{a}\cos t~~~\mbox{and}~~~
x=-\frac{a}{2\epsilon\cos t}-\frac{2\epsilon\cos t}{a}
\end{equation}
These solutions must match the solution (\ref{eqn:38}) and this
matching should give the time $t$. The first solution of
(\ref{eqn:44}) is of $O(\epsilon)$ and cannot ever match the
solution (\ref{eqn:38}) or (\ref{eqn:36}) which is of $O(1)$. On the other hand
the second solution of (\ref{eqn:44}) is of $O(1/\epsilon)$, i.e. a large
number, and this can match the solution (\ref{eqn:38}) if $x_0$ is
large. In fact in Fig.8 we see that there are solutions with $y=0$
for $|x|>8$.

In conclusion, similarly to the case considered in EKC, for
$\epsilon$ small the nodal lines leave an empty central domain,
delimited by hyperbola-like boundaries.
\begin{figure}\label{nd22}
\centering
\includegraphics[scale=0.8]{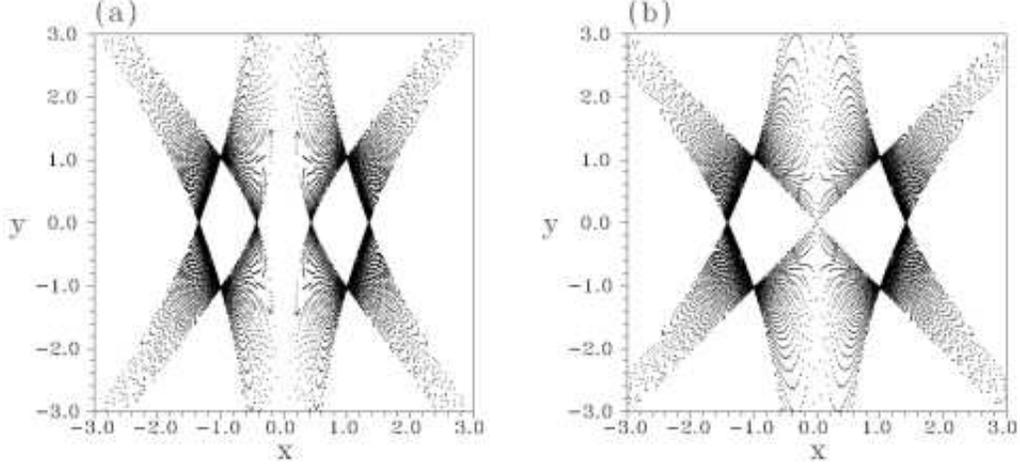}
\caption{\small Nodal lines of the wavefunction
$\Psi=\Psi_{00}+a\Psi_{20}+b\Psi_{11}$ when $c=\sqrt{2}/2$ and (a)
$a=1.23$, $b=1.15$, (b) $a=1$, $b=1.15$.}
\end{figure}
\begin{figure}\label{nd30}
\centering
\includegraphics[scale=0.7]{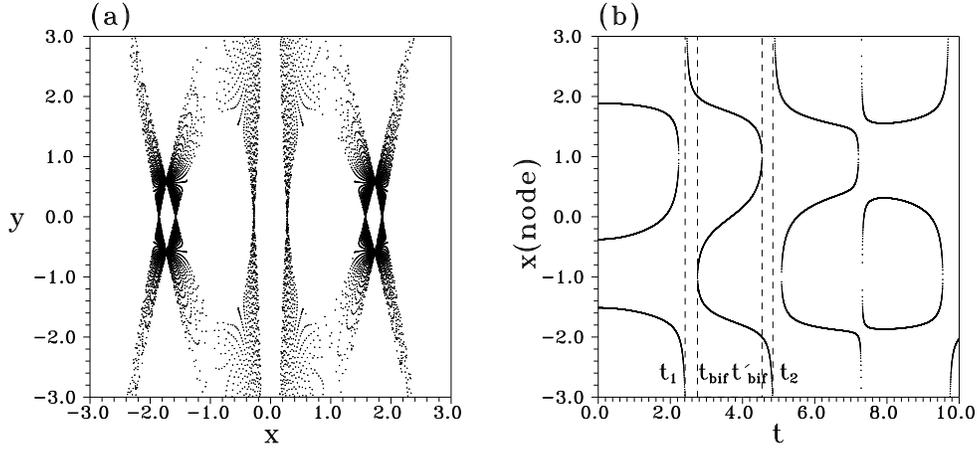}
\caption{\small (a) Nodal lines of the wavefunction
$\Psi=\Psi_{00}+a\Psi_{30}+b\Psi_{11}$ when $a=1.23$, $b=1.15$,
and $c=\sqrt{2}/2$. (b) The time evolution of $x_0(t)$ for either
one or the three solutions of (\ref{eqn:28}).}
\end{figure}

\subsubsection{Case $\Psi=\Psi_{00}+a\Psi_{20}+b\Psi_{11}$}

In the same way as above, the real and imaginary parts of $\Psi=0$
yield
\begin{eqnarray}\label{eqn:18}
&~&1+a\left(x^2-1\right)\cos 2t+bx\sqrt{c}y\cos
\left(1+c\right)t=0\nonumber\\
&~&a\left(x^2-1\right)\sin 2t +bx\sqrt{c}y\sin
\left(1+c\right)t=0~~.
\end{eqnarray}
Multiplying the second equation by $\cot\left(1+c\right)t$ and
subtracting from the first we find
\begin{equation}
\sin\left(1+c\right)t+a\left(x^2-1\right)\sin \left(c-1\right)t=0
\end{equation}
hence
\begin{equation}\label{eqn:21}
x^2=1-\frac{\sin\left(1+c\right)t}{a\sin \left(c-1\right)t},~~~
y=(x\sqrt{c})^{-1}\bigg[\frac{a\left(x^2-1\right)\sin 2t}{b\sin
\left(c+1\right)t}= \frac{\sin 2t}{b\sin
\left(c-1\right)t}\bigg]~~.
\end{equation}
Equations (\ref{eqn:21}) give two nodal points with opposite $x$
and $y$ whenever the restriction $x^2\geq 0$ is satisfied, and no
nodal points when it is not. Thus, the nodal points exist only in
particular time intervals.

In general the nodal lines enclose an empty region near the origin
(Fig.9a). However in the particular case $a=1$ the nodal lines
reach the center $x=y=0$ (Fig.9b). In fact, $y=0$ if $\sin 2t=0$,
i.e. $t=0, \pi/2, \pi...$ The corresponding values of $x^2$ are :
For $t=0$, $x^2=1-\frac{1+c}{a\left(c-1\right)}$, for $t=\pi/2$,
$x^2=1+\frac{1}{a}$,  for $t=\pi$, $x^2=1-\frac{1}{a}$ etc. The
last solution exists only if $a\geq 1$, and it yields $x=0$ if
$a=1$ exactly. On the other hand the solution $x^2=1+\frac{1}{a}$
always exists if $a>0$, meaning that the nodal lines intersect the
$x-$axis at the points $x=\pm(1+1/a)^{1/2}$ (Figs.9a,b).

\subsubsection{Case $\Psi=\Psi_{00}+a\Psi_{30}+b\Psi_{11}$}

In this case the real and imaginary parts of $\Psi=0$ yield
\begin{eqnarray}\label{eqn:25}
&~&1+a\left(x^3-3x\right)\cos 3t+bx\sqrt{c}y\cos
\left(1+c\right)t=0\nonumber\\
&~&a\left(x^3-3x\right)\sin 3t +bx\sqrt{c}y\sin
\left(1+c\right)t=0~~.
\end{eqnarray}
Multiplying the first equation by $\sin \left(1+c\right)t$ and the
second equation by $\cos \left(1+c\right)t$ and subtracting we find
\begin{equation}
\sin\left(1+c\right)t+a\left(x^3-3x\right)\sin \left(c-2\right)t=0
\end{equation}
hence
\begin{equation}\label{eqn:28}
x^3-3x+\frac{\sin\left(1+c\right)t}{a\sin
\left(c-2\right)t}=0,~~~y=\frac{\sin 3t}{b\sqrt{c}x\sin
\left(c-2\right)t}
\end{equation}
The third degree equation has three real roots if
\begin{equation}\label{eqn:29}
\frac{\sin^2\left(1+c\right)t}{4a^2\sin^2 \left(c-2\right)t}\leq
1,
\end{equation}
otherwise only one root is real.

The nodal lines in this case have the form of Fig.10a, leaving again
empty central domains in the configuration space. We have
three nodal points if the inequality (\ref{eqn:29}) is satisfied,
and one nodal point otherwise. In fact, there are distinct time
intervals within which one of the nodal points comes from or goes
to infinity (Fig.10b). For example, after the time $t_1=2.43$, one
nodal point (point 1) starts approaching the central region from
$x\rightarrow\infty$. A little later (at $t=t_{bif}=2.75$),
a pair of nodal points (2 and 3) emerge at $x\simeq -1$.
At the time $t=t_{bif}'=4.54$, point 2 joins point 1 nearly at $x=1$,
and after $t=t_{bif}'$ these two points disappear, while point 3 tends
to $x\rightarrow -\infty$ at $t=t_2=4.85$. Similar phenomena take place
at subsequent intervals of time. The times $t_{bif},t'_{bif}$, etc. are
called `bifurcation times'. Such
bifurcations are important for the level of chaos of the trajectories
approaching the nodal points, because close to a bifurcation time
the speed of the bifurcating nodal points (e.g. $\dot{x}_0$) which
enters into the estimates of local Lyapunov characteristic numbers
(Eq.(\ref{xiovx0})), is large (see numerical simulations below).

\subsection{The degree of chaos for ensembles of chaotic trajectories}

As a first example, we consider orbits in the $\psi$-field
$\Psi=\Psi_{00}+a\Psi_{20}+b\Psi_{11}$ (subsection III C 2) when
$a=1.23$, $b=1.15$, $c=\sqrt{2}/2$. The nodal points in this field
appear in pairs, within some time intervals. The X-points
are calculated by a Newton-Raphson method with a precision tolerance of
$10^{-14}$, loading (\ref{1v3}) as initial guess values of $(u_X,v_X)$.
Fixing the frame of reference on one of the nodal points, two X-points
are found numerically. One X-point is close to the considered nodal point
and the other is far from it. Nevertheless, the distant X-point is irrelevant
to the dynamics, because in that case we have (for, say, the nodal point
$(x_{01},y_{01})$) $|\vec{V}-\vec{V}_{02}|=|\vec{V}_{01}-\vec{V}_{02}|$
$=2|\vec{V}_{01}|=2|\vec{V}|$ (since $\vec{V}_{02}=-\vec{V}_{01}$),
implying $|\vec{V}-\vec{V}_{02}|/|\vec{V}|=2$, i.e.,
a violation of the condition (\ref{adcon1}). Thus, in the numerical
calculations we only take into account approaches to the X-points found
in the vicinity of each nodal point in its own frame of reference.
\begin{figure}\label{orb22}
\centering
\includegraphics[scale=0.8]{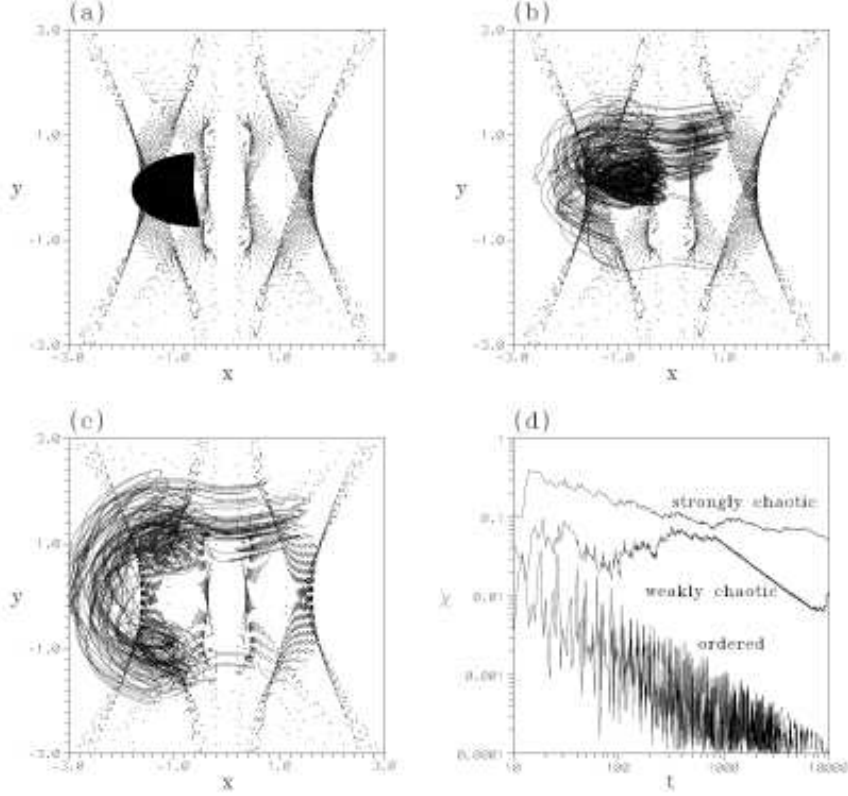}
\caption{\small Examples of quantum trajectories in the $\psi-$
field $\Psi=\Psi_{00}+a\Psi_{20}+b\Psi_{11}$ when $a=1.23$,
$b=1.15$, and $c=\sqrt{2}/2$. (a) An ordered orbit (initial
conditions $x(0)=-1.5$, $y(0)=0.1275$, (b) a weakly chaotic orbit
(initial conditions $x(0)=0.850901842117$, $y(0)=1.191571712494$,
and (c) a strongly chaotic orbit (initial conditions
$x(0)=-1.231356695294$, $y(0)=0.840584903955$. In all three panels
the orbits are plotted up to $t=1000$. (d) Time evolution of the
`finite time Lyapunov characteristic number' $\chi(t)$ for the
three orbits. }
\end{figure}

For any trajectory with initial conditions $x(0),y(0)$ there is a
centrally symmetric orbit with initial condition $-x(0),-y(0)$.
Figure 11 shows three orbits: regular (Fig.11a), weakly chaotic
(Fig.11b), and strongly chaotic (Fig.11c). The degree of chaos
is measured by the quantity
\begin{equation}\label{chi}
\chi(t)={1\over t}\ln{\xi(t)\over\xi(0)}
\end{equation}
where $\xi(t)$ is the length, at time $t$, of a deviation
$\vec{\xi}=(\Delta x,\Delta y)$ from an orbit $\bigg(x(t),y(t)\bigg)$,
calculated by the variational equations of motion. This is called
`finite time Lyapunov characteristic number' and the limit
$\lim_{t\rightarrow\infty}\chi(t)$ yields the Lyapunov
characteristic number of an orbit. In the case of the regular
orbit, the quantity $\chi(t)$ (Fig.11d) decreases as a power law
$\chi(t)\sim t^{-1}$. In the case of the strongly chaotic orbit,
after some transient time the quantity $\chi(t)$ decreases slowly
and it tends to stabilize to a value $\chi(t)\simeq 5\times 10^{-2}$
at $t=10^4$. On the other hand, in the case of the weakly chaotic
orbit, there is a temporary stabilization of $\chi(t)$ up to $t=10^3$,
followed, however, by a $t^{-1}$ decrease up to $t=7\times 10^3$.
Beyond this time $\chi(t)$ increases again up to the value
$\chi=10^{-2}$ at $t=10^4$, showing no signs of stabilization.

In general, weakly chaotic orbits exhibit long transient intervals
in which $\chi$ fluctuates around values typically one order of
magnitude smaller than the stabilization value of the strongly
chaotic orbits. A careful inspection shows that in most cases this
behavior of the weakly chaotic orbits can be characterized as a
`stickiness' phenomenon (see e.g. \cite{cont2002}), namely the
orbits behave essentially as regular in a transient time interval.
In Fig.11b this tendency is observed for the weakly chaotic orbit,
which, besides the chaotic oscillations, shows a domain of
enhanced density similar to the domain filled by the regular orbit
of Fig.11a.

Also, the difference in the evolution of $\chi(t)$ for
the three orbits is related to their frequency of encounters with nodal
point - X-point complexes. The background points in Figs.11a,b,c
show the distribution of the X-points in the configuration
space, which remains practically unaltered after a time $t=1000$.
The X-points occupy domains similar to those occupied by the nodal
points (Fig.9a). Setting an upper threshold distance $d_{max}$,
and splitting the time evolution of the orbits into time segments
of width $\Delta t$, we may count the number of time windows
within which a trajectory approached the X-point at a minimum
distance $d\leq d_{max}$. In the numerical calculations we set
$d_{max}=0.2$, and $\Delta t=0.1$, and find a number of
approaches, up to a time $t=10^4$, equal to
$N(d\leq 0.2)=0$ for the regular orbit, $N(d\leq 0.2)=106$ for the
weakly chaotic orbit and $N(d\leq 0.2)=389$ for the strongly
chaotic orbit. One can check that the ratios of the values of
$N(d<d_{max})$ of the three orbits remain practically invariant if
another choice of $d_{max}$ is made, provided that $d_{max}$ is
bounded (e.g. $d_{max}$ does not exceed unity).

In the system $\Psi=\Psi_{00}+a\Psi_{30}+b\Psi_{11}$ (subsection III
C 3, one or three nodal points) we also find regular, chaotic and
weakly chaotic orbits. In the system of subsection III C 1 we find
orbits similar to those of EKC as long as the parameter $\epsilon$
is small.
\begin{figure}\label{chigrd}
\centering
\includegraphics[scale=0.8]{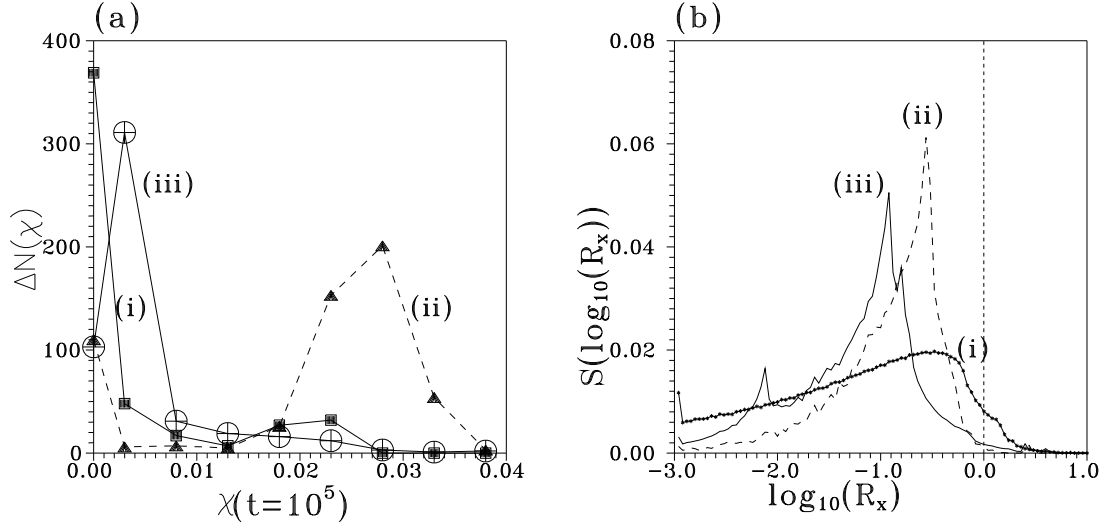}
\caption{\small Histograms (line diagrams) of the number of orbits
$\Delta N$ of which the finite time Lyapunov characteristic number at $t=10^5$
is in the interval $[\chi-0.0025,\chi+0.0025)$, where $\chi$ is
the value shown in the abscissa. The orbits result from 500
initial conditions taken randomly in the box $-1.5\leq x\leq 1.5$,
$0\leq y\leq 1.5$. The solid line with squares corresponds to the EKC
model (i) $\Psi=\Psi_{00}+a\Psi_{10}+b\Psi_{11}$ (one nodal point),
the dashed line to the model (ii) $\Psi=\Psi_{00}+a\Psi_{20}+b\Psi_{11}$
(two nodal points), and the solid line with crossed circles to the model
(iii) $\Psi=\Psi_{00}+a\Psi_{30}+b\Psi_{11}$ (three or one nodal point).
In all three cases $a=1.23$, $b=1.15$, $c=\sqrt{2}/2$. (b) Histogram of the
values of $log_{10}R_x$ (nodal point - X-point distance) when all the nodal
point - X-point complexes are calculated at the times $t=n\times
10^{-1}$, $n=1,...,10^5$. The dotted solid line refers to the EKC
model (i), the dashed line to the model (ii), and the solid line to the
model (iii), with parameters as in (a). }
\end{figure}

Figure 12 shows now the main result. Having run 500 orbits with
initial conditions taken randomly in the box $-1.5\leq x\leq 1.5$,
$0\leq y\leq 1.5$ in the three systems (i)
$\Psi=\Psi_{00}+a\Psi_{10}+b\Psi_{11}$ (EKC, one nodal point),
(ii) $\Psi=\Psi_{00}+a\Psi_{20}+b\Psi_{11}$ (two nodal points) and
(iii) $\Psi=\Psi_{00}+a\Psi_{30}+b\Psi_{11}$ (three or one nodal
points), Fig.12a compares the distributions of the finite time
Lyapunov characteristic numbers $\chi(t=10^5)$ for the three
ensembles of orbits. The time $t=10^5$
is long enough to extinguish transient effects of $\chi(t)$ for
most orbits. It is immediately clear that model (ii) yields a
significantly larger degree of chaos than models (i) and (iii).
Model (i) yields a bimodal distribution, with a large proportion
of regular orbits ($\chi=0$) and also a local maximum of the
distribution of the chaotic orbits at $\chi\approx 0.025$. Model
(iii), on the other hand, has a small number of perfectly regular
orbits, but the main bulk of its chaotic orbits is also in rather
small values of $\chi$ ($\chi<0.01$).

In the systems (ii) and (iii) the double or triple nodal points appear
only in certain time intervals. In a total time $t=10^4$,
divided in segments $\Delta t=10^{-1}$, the total number of nodal point
- X-point complexes detected in a box $(x,y)\in [-3,3]\times [-3,3]$
are $8.4\times 10^4$ for the system (i), $5.1\times 10^4$ for the system
(ii), and $1.47\times 10^5$ for the system (iii). Thus, chaos appears
{\it less pronounced} precisely in the system exhibiting the largest
number of nodal point - X-point complexes, i.e. the system (ii).
This phenomenon can be understood if we take into account the
theory of subsection II D, and in particular the fact that the values
of the local Lyapunov characteristic number have a
$O(V^{-1})$ dependence on the speed of the nodal point (Eq.(\ref{xiovx0})),
or, equivalently, a $O(R_X)$ dependence on the nodal point - X-point distance.
Plotting the histograms of the values of $R_X$ for all three systems (Fig.12b)
renders immediately clear that in the system (ii) (two nodal points) the main
bulk of the histogram is at values of $R_X$ larger than in both the systems
(i) and (iii) (one or three nodal points), i.e. the system (ii) has the
more effective chaotic scatterers (complexes) from all three systems.
In the system (ii), $R_X$ is mainly distributed over the range
$0.1\leq R_X\leq 1$ (with a mean $<R_x>\simeq 0.2$). Thus, both the
condition of validity of the adiabatic approximation ($R_X<1$, vertical
dashed line in Fig.12b) and the
requirements for effective chaotic scattering ($R_X$ large) are fulfilled.
In the system (iii), the speed of bifurcating nodal points is large near the
`bifurcation times' $t_{bf}$ (see Fig.10b and the relevant discussion), and
this reduces the effectiveness of the chaotic
scattering.  In the case of the system (i), there is a significant
percentage of non-effective complexes ($R_X<10^{-2}$) or of complexes
violating the condition of adiabaticity (i.e. with $R_X>1$). In this
system we thus find less chaos than in the system (ii), and also a
large number of perfectly regular orbits.
\begin{figure}\label{gridall}
\centering
\includegraphics[scale=0.8]{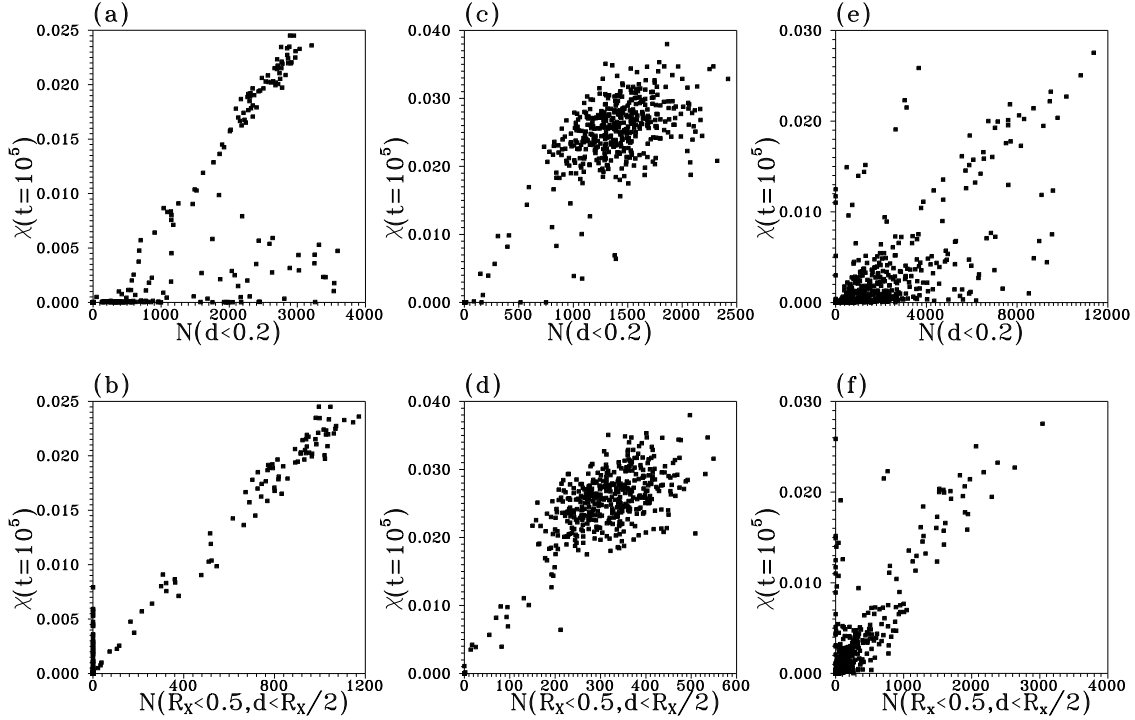}
\caption{\small The finite time Lyapunov characteristic number
$\chi$ at $t=10^5$ as a function of the number $N(d<0.2)$ of
approaches of an orbit to the X-point at a distance smaller or
equal to 0.2 for the systems (i),(ii) and (iii), shown in (a), (c)
and (e) respectively. For each system, 500 initial conditions are
taken randomly in the box $-1.5\leq x\leq 1.5$, $0\leq y\leq 1.5$.
The finite time Lyapunov characteristic number $\chi$ versus the
corrected index $N(R_X\leq 0.5,d\leq R_X/2)$ (see text) is shown
in (b), (d) and (f) respectively.}
\end{figure}

The value of $\chi$ for particular orbits is in general an increasing
function of the number of encounters with nodal point - X-point complexes,
but this relation presents considerable scatter and also noticeable
exceptions.  In Figs.13a,c,e the number of consecutive approaches
of a trajectory to a nodal point X-point complex are estimated by the
index $N(d\leq 0.2)$ (number of approaches at a distance $d\leq 0.2$;
similar results are found if $N(d\leq 0.5)$ is used instead).
The tendency of $\chi$ to increase with $N$ is clear
in all three systems. The exceptions refer to orbits in the lower right
part of each of Figs.13a,c,e. These exceptions disappear, however, if
we use a corrected index for the number of encounters, by the
requirement that an encounter is only counted provided that the
size $R_X$ of the complex during it is not very small. Figures 13b,d,f
show the scaling of $\chi$ versus the corrected index
$N(R_X\leq 0.5,d\leq R_X/2)$, for the same orbits, i.e. we count only
the encounter events in which
the distance $R_X$ is not larger than twice the distance $d$
at which the orbits have the closest local approach to a nodal
point X-point complex (the factor two is rather arbitrary; in
general we can set $R_X>O(d)$). The second condition, $R_X\leq
0.5$, ensures that only complexes being well within the regime of
validity of the adiabatic approximation are selected (this is also
arbitrary; any limit $R_X<O(1)$ can be used). These extra
conditions immediately yield a lower number of recorded events for
all the orbits than by the index $N(d<0.2)$. The main effect
however is that all the exceptions of Figs.13a,c,e disappear
by utilizing the corrected index, thus yielding a better correlation
of $\chi$ with $N(R_X\leq 0.5,d\leq R_X/2)$.

\section{Conclusions}
We developed the general theory of motion in the vicinity of a moving
2D `quantum vortex', i.e. a nodal point of the wavefunction, in the
trajectory (Bohmian) approach of the quantum flow, and we discussed
the origin and quantification of chaos for the Bohmian trajectories.
Our main findings can be summarized as follows:

1) The flow in the vicinity of a moving nodal point is non-autonomous,
but under suitable `adiabatic' conditions it can be treated as nearly
autonomous. Two necessary and sufficient conditions are found:
a) the equations of motion must be taken in a moving frame of reference
centered at the nodal point, and b) the latter's velocity must be large
in the rest frame.

2) Developing an arbitrary wavefunction $\psi$ up to terms of second
degree with respect to the distance from the nodal point, we demonstrate
that the appearance of nodal point - X-point complexes is a generic
feature of the configuration space. There are two stable and two
unstable manifolds emanating from the X-point associated to each
nodal point. One of these manifolds continues as a spiral approaching
the nodal point, while two other form very narrow channels allowing
communication with the interior of the complex. The nodal point undergoes
consecutive Hopf bifurcations. Whenever a Hopf bifurcation takes place
a limit cycle is formed around the nodal point for transient time
intervals. As a consequence of all these facts, it is shown that
most trajectories do not penetrate deeply into the complex.

3) On the other hand, the chaotic orbits are scattered by the complex
via encounters of `type I' (forming a loop around the complex) or of
`type II' (no loop). A theoretical estimate is given of the local Lyapunov
characteristic numbers in separate encounter events. The local Lyapunov
characteristic number scales as an inverse power of the speed $V$ of the nodal
point and of the distance of the scattered trajectory from the X-point's stable
manifold (impact parameter) far from the complex.
The size of the complex (distance $R_X$ of the X-point from the nodal
point) scales as $R_x\sim V^{-1}$. The chaotic scattering is most effective
when the speed of the nodal point is relatively small, or $R_X$ is
large. But $R_X$ is also limited by the extra condition of adiabaticity ($R_X<1$).
Numerically, we find most effective chaotic scattering events taking
place in the range $0.01\leq R_X\leq 1$, $R_X\sim 0.1$ being an optimal
value.

4) We provide numerical examples of the loci occupied by the nodal
points and the X-points in different examples of superposition of
a number of eigenstates in a 2D harmonic potential model. In particular,
we examine three models with (i) one, (ii) two, or (iii) three nodal points,
and identify the domains of each system devoid of nodal points.
The trajectories having no overlap with the domains of nodal points
turn to be regular. There are also weakly chaotic trajectories, exhibiting
stickiness phenomena, and strongly chaotic orbits having a significant
overlap with the domains of nodal points. The system with the smaller
number of complexes (system (ii)) turns to have the largest degree of
chaos. This is explained by examining carefully the properties of
the complexes and demonstrating (on the basis of the theory of section
II) that the most effective chaotic scattering events are produced in
the case of the system (ii).

5) The `finite time Lyapunov characteristic numbers $\chi(t)$ of the trajectories
have a nearly linear correlation with the number of encounters with nodal
point - X-point complexes, but with considerable scatter.
The scatter is reduced, and most exceptions disappear when only `effective'
events are counted. The effectiveness criterion takes into account
the requirement that the trajectory approaches the X-point at moments
when the complex is relatively large, implying that the chaotic scattering
is strong.

\begin{acknowledgments}

C. Kalapotharakos was supported by the Research Committee of the
Academy of Athens. We thank two anonymous referees for their remarks.

\end{acknowledgments}

\appendix

\section{Equations of motion for a generic $\psi-$field.}

The first of Eqs.(\ref{eqmo2}) is obtained by multiplying the
first and the second of Eqs.(\ref{eqmoexp}) by $u$ and $v$
respectively and adding the results. We then find
$u(du/dt)+v(dv/dt)=RdR/dt$. Similarly, the second of
Eqs.(\ref{eqmo2}) is obtained by multiplying the first and the
second of Eqs.(\ref{eqmoexp}) by $v$ and $u$ respectively and
subtracting the results. We then find
$v(du/dt)-u(dv/dt)=Rd\phi/dt$. After these operations, the values
of the coefficients $c_2,c_3,d_0$ and $d_1$ of Eq.(\ref{eqmo2})
are readily evaluated. Together with the conditions $a_{02}=-a_{20}$,
$b_{02}=-b_{20}$, the average value of the coefficient $f_3$:
$$
<f_3>(a_{ij},b_{ij},V_x,V_y)={1\over 2\pi}
\int_{0}^{2\pi}\bigg({c_3\over d_0}-{c_2d_1\over d_0^2}\bigg)d\phi
$$
takes the form:
$$
<f_3>={1\over
4(a_{10}b_{01}-a_{01}b_{10})^2}\times\bigg[V_x\bigg(2a_{01}a_{10}^2b_{02}+a_{01}^2a_{11}b_{10}
-a_{10}^2a_{11}b_{10}-2a_{10}a_{02}a_{01}b_{10}+a_{11}b_{01}^2b_{10}
$$
$$
+2a_{10}b_{01}b_{02}b_{10}-2a_{02}b_{01}b_{10}^2-a_{11}b_{10}^3-a_{01}^2a_{10}b{11}+a_{10}^3b_{11}
-a_{10}b_{01}^2b_{11} +a_{10}b_{10}^2b_{11}\bigg)
$$
$$
-V_y\bigg(2a_{01}a_{02}a_{10}b_{01}-2a_{10}a_{01}^2b_{02}+a_{10}^2a_{11}b_{01}
-a_{01}^2a_{11}b_{01}+a_{11}b_{10}^2b_{01}
$$
$$
+2a_{02}b_{10}b_{01}^2-2a_{01}b_{10}b_{02}b_{01}-a_{11}b_{01}^3-a_{10}^2a_{01}b_{11}+a_{01}^3b_{11}
-a_{01}b_{10}^2b_{11} +a_{01}b_{01}^2b_{11}\bigg)
$$
$$
+(V_x^2-V_y^2)\bigg(a_{01}^3a_{10}+a_{01}a_{10}^3+a_{01}a_{10}b_{01}^2
+a_{01}^2b_{01}b_{10}+a_{10}^2b_{01}b_{10}+b_{01}^3b_{10}
+a_{01}a_{10}b_{10}^2+b_{01}b_{10}^3\bigg)
$$
$$
+V_xV_y\bigg(a_{01}^4-a_{10}^4+2a_{01}^2b_{01}^2+b_{01}^4-b_{10}^4-2a_{10}^2b_{10}^2\bigg)~~.
$$
\\
Since all the terms in the above expression have either $V_x$ or
$V_y$ as a coefficient, it follows that $<f_3>=0$ if $V_x=V_y=0$,
i.e. the nodal point is a center in the rest frame, and an
attractor or repellor in any other moving frame of reference.

\section{Local growth of the deviations $\xi(t)$
in a trajectory - nodal point - X-point scattering event}

Referring to the model (\ref{eqrot2}) of subsection II D, let
$T(C)$ be the time required for an orbit to traverse the complex
along one of the integral curves given by (Eq.\ref{ccon}). For
simplicity (and without loss of generality) we consider the case
$\dot{x}_0>0$ and identify $T(C)$ to the time needed for an orbit
starting on a curve $C$, at $u=1$, until the orbit crosses $u=-1$,
namely:
\begin{equation}\label{tc}
T(C)=2\int_{v_1(C)}^{v_0(C)}{Ce^{-2\dot{x}_0v}dv\over\sqrt{Ce^{-2\dot{x}_0v}-v^2}}
\end{equation}
where  $v_0$ and $v_1$ correspond to the $v-$values satisfying
\begin{equation}\label{v0v1}
e^{2\dot{x}_0v_0}v_0^2=C=e^{2\dot{x}_0v_1}(1+v_1^2)
\end{equation}
i.e. the values of $v$ on the curve $C$ for $u=0$ and $u=u_1=\pm
1$ respectively. The X-point is located at
\begin{equation}\label{vx}
u_x=0,~~v_x=-{1\over\dot{x}_0}
\end{equation}
and its asymptotic curves have the C-value
\begin{equation}\label{cx}
C=C_x={1\over e^2\dot{x}_0^2}~~.
\end{equation}
When $\dot{x}_0$ is large the asymptotic curves become nearly
horizontal a little further from the X-point, i.e. $|v_1(C_x)|$ is
small. The same holds true for nearby curves with $C\approx C_x$.
The value of $v_1$ can then be found approximately by expanding
$\ln(C)=2\dot{x}_0v_1+\ln(1+v_1^2)$ to second order in $v_1$,
yielding
\begin{equation}\label{v1c}
v_1(C)\simeq {\ln C\over \dot{x}_0+\sqrt{\dot{x}_0^2+\ln C}}~~.
\end{equation}
In view of (\ref{cx}) we have $|\ln C|\sim
2\ln|\dot{x}_0|<<|\dot{x}_0|$. Thus, setting
$v_1(C)\approx\ln(C)/(2\dot{x}_0)$ is nearly always a sufficient
approximation.

Consider now the orbits on two neighboring integral curves
$C,C+\Delta C$ such that $T(C)>T(C+\Delta C)=T(C)+\Delta T$. Far
from the complex the velocity of the orbits is $|\vec{V}|\approx
-\dot{x}_0$. It follows that at the time $t=T(C)$ the two orbits
are a distance $\sim |\dot{x}_0\Delta T|$ apart. Thus the initial
deviation $dv_1\equiv \xi_0$ has grown to
$\xi\approx\xi_0+|\dot{x}_0\Delta T|$. But $\Delta T=(dT/dC)\Delta
C$ and by virtue of (\ref{v1c}) $\Delta C\simeq
2C\dot{x}_0dv_1=2C\dot{x}_0\xi_0$. Thus, the final value of the
deviation can be estimated as:
\begin{equation}\label{xifin}
\xi\approx \xi_0\bigg(1+2\dot{x}_0^2\bigg{|}{dT\over dC}\bigg{|}C\bigg)~~.
\end{equation}
Equation (\ref{xifin}) states that the growth of deviations is proportional
to the differential rate of description of the integral curves passing
close to the X-point. If $u>1$ the velocity stabilizes to $v\simeq -\dot{x}_0$
along all the integral curves. This explains the stabilization of $\xi$ in
Fig.2b. Furthermore, $|dT/dC|$ increases as $C$ tends to $C_x$, since
$\lim_{C\rightarrow C_x}|dT/dC|=\infty$. The singular behavior at
$v=v_s$, $C=C_x$, corresponds to the two peaks of Fig.3a.

The following is an explicit calculation of the value of $\xi$ reached
asymptotically for type II orbits (the calculation is similar for type I
orbits). Expanding $v_0$ as $v_0=v_x -\delta v_0 = -1/\dot{x}_0 - \delta v_0$,
we find from Eq.(\ref{ccon}) that the first order variations cancel exactly.
The second order variations yield:
\begin{equation}\label{v0down}
v_0(C)\simeq -{1\over\dot{x}_0}(1+|Ce^2\dot{x}_0^2-1|^{1/2})
\end{equation}
(a similar calculation for type I events yields that the
separatrix intersects the $v-$axis at the positive $v-$value
$v_x^{up}=a_s/\dot{x}_0$, $a_s=0.278464...$ being the root of
$a_s+\ln a_s+1=0$; for type I curves the upper intersection with
the $v-$ axis is found through first variations of
Eq.(\ref{ccon}), namely $v_0^{up}(C) =
(1/\dot{x}_0)(0.278464+0.108906|Ce^2\dot{x}_0^2-1|)$).

The asymptotic behavior of the integral $T(C)$ with respect to $C$
can now be found by isolating the singularity of the integrand at
$v=v_0(C)$
\begin{equation}\label{tcexp}
T(C)=2\int_{|v_1(C)-v_0(C)|}^{0}{Ce^{-2\dot{x}_0v_0}(1-2\dot{x}_0\Delta
+2\dot{x}_0^2\Delta^2+...)d\Delta
\over\sqrt{(2\dot{x}_0Ce^{-2\dot{x}_0v_0}+2v_0)\Delta
+(2C\dot{x}_0^2e^{-2\dot{x}_0v_0}-1)\Delta^2+...}}
\end{equation}
where $\Delta\equiv |v-v_0(C)|$. The exact value of the lower
limit used in this integral does not really matter in the
calculation of $dT/dC$, since the leading contribution to $T(C)$
comes from the parts of the orbits close to the X-point, i.e. for
$\Delta$ small; the lower limit can actually be substituted by a
value $\Delta_{max}\sim 1/\dot{x}_0$ ensuring that the truncated
expansion in the square root is a sufficient approximation. On the
other hand, it is necessary to retain $O(\Delta^2)$ terms in the
expansion within the square root of Eq.(\ref{tcexp}), because the
$O(\Delta)$ term becomes very small as $C$ tends to $C_x$, while
the second order term is always of order unity. The upper limit of
the integral (\ref{tcexp}) yields then logarithmic terms:
\begin{equation}\label{tcasym}
T(C)\approx
2Ce^{-2\dot{x}_0v_0}(I_1-2\dot{x}_0I_2+2\dot{x}_0^2I_3)
\end{equation}
$$\mbox{with}~~~~~~~~
I_1=ln[8(\dot{x}_0Ce^{-2\dot{x}_0v_0}+v_0)](2C\dot{x}_0^2e^{-2\dot{x}_0v_0}-1)^{-1/2}
$$
$$
I_2=-(\dot{x}_0Ce^{-2\dot{x}_0v_0}+v_0)ln[8(\dot{x}_0Ce^{-2\dot{x}_0v_0}+v_0)]
(2C\dot{x}_0^2e^{-2\dot{x}_0v_0}-1)^{-3/2}
$$
$$
I_3=(3/2)(\dot{x}_0Ce^{-2\dot{x}_0v_0}+v_0)^2ln[8(\dot{x}_0Ce^{-2\dot{x}_0v_0}+v_0)]
(2C\dot{x}_0^2e^{-2\dot{x}_0v_0}-1)^{-5/2}~.
$$
As $C\rightarrow C_x$,
$(\dot{x}_0Ce^{-2\dot{x}_0v_0}+v_0)\rightarrow 0$, $I_1$ becomes
singular while $I_2$ and $I_3$ are finite. Taking into account
that $Ce^{-2\dot{x}_0v_0}=v_0^2$, we then find, to the leading
order,
\begin{equation}\label{dcdt}
\dot{x}_0^2C|{dT(C)\over dC}|\propto \Bigg({\dot{x}_0^2v_0\over
\dot{x}_0v_0+1}\Bigg)\Bigg({v_0^2\dot{x}_0\over(\dot{x}_0v_0+1)\sqrt{2\dot{x}_0^2v_0^2-1}}+\ldots\Bigg)
\end{equation}
or, using Eq.(\ref{xifin}) and $v_0=-1/\dot{x}_0-\delta v_0$,
\begin{equation}\label{xiovxv0}
{\xi\over\xi_0}\sim {1\over \dot{x}_0^2\delta v_0^2}+...
\end{equation}
However, in view of Eqs.(\ref{v0down}), (\ref{v1c}) and (\ref{cx})
we have $\dot{x}_0\delta v_0^2\propto \delta v_1$, thus
\begin{equation}\label{xiovx0n}
{\xi\over\xi_0}\sim {1\over \dot{x}_0\delta v_1}+...
\end{equation}
that is we obtain the power-law estimate of Eq.(\ref{xiovx0}).

\bibliography{EKC3quantph}

\begin{thebibliography}{49}
\expandafter\ifx\csname natexlab\endcsname\relax\def\natexlab#1{#1}\fi
\expandafter\ifx\csname bibnamefont\endcsname\relax
  \def\bibnamefont#1{#1}\fi
\expandafter\ifx\csname bibfnamefont\endcsname\relax
  \def\bibfnamefont#1{#1}\fi
\expandafter\ifx\csname citenamefont\endcsname\relax
  \def\citenamefont#1{#1}\fi
\expandafter\ifx\csname url\endcsname\relax
  \def\url#1{\texttt{#1}}\fi
\expandafter\ifx\csname urlprefix\endcsname\relax\def\urlprefix{URL }\fi
\providecommand{\bibinfo}[2]{#2}
\providecommand{\eprint}[2][]{\url{#2}}

\bibitem[{\citenamefont{Dirac}(1931)}]{dirac1931}
\bibinfo{author}{\bibfnamefont{P.~A.~M.} \bibnamefont{Dirac}},
  \bibinfo{journal}{Proc. R. Soc. A} \textbf{\bibinfo{volume}{133}},
  \bibinfo{pages}{60} (\bibinfo{year}{1931}).

\bibitem[{\citenamefont{Hirschfelder
  et~al.}(1974{\natexlab{a}})\citenamefont{Hirschfelder, Goebel, and
  Bruch}}]{hgb1974}
\bibinfo{author}{\bibfnamefont{J.}~\bibnamefont{Hirschfelder}},
  \bibinfo{author}{\bibfnamefont{C.~J.} \bibnamefont{Goebel}},
  \bibnamefont{and} \bibinfo{author}{\bibfnamefont{L.~W.} \bibnamefont{Bruch}},
  \bibinfo{journal}{J. Chem. Phys.} \textbf{\bibinfo{volume}{61}},
  \bibinfo{pages}{5456} (\bibinfo{year}{1974}{\natexlab{a}}).

\bibitem[{\citenamefont{Hirschfelder
  et~al.}(1974{\natexlab{b}})\citenamefont{Hirschfelder, Christoph, and
  Palke}}]{hcp1974}
\bibinfo{author}{\bibfnamefont{J.}~\bibnamefont{Hirschfelder}},
  \bibinfo{author}{\bibfnamefont{A.~C.} \bibnamefont{Christoph}},
  \bibnamefont{and} \bibinfo{author}{\bibfnamefont{W.~E.} \bibnamefont{Palke}},
  \bibinfo{journal}{J. Chem. Phys.} \textbf{\bibinfo{volume}{61}},
  \bibinfo{pages}{5435} (\bibinfo{year}{1974}{\natexlab{b}}).

\bibitem[{\citenamefont{Skodje et~al.}(1989)\citenamefont{Skodje, Rohrs, and
  VanBuskirk}}]{srv1989}
\bibinfo{author}{\bibfnamefont{R.~T.} \bibnamefont{Skodje}},
  \bibinfo{author}{\bibfnamefont{H.~W.} \bibnamefont{Rohrs}}, \bibnamefont{and}
  \bibinfo{author}{\bibfnamefont{J.}~\bibnamefont{VanBuskirk}},
  \bibinfo{journal}{Phys. Rev. A} \textbf{\bibinfo{volume}{40}},
  \bibinfo{pages}{2894} (\bibinfo{year}{1989}).

\bibitem[{\citenamefont{Lopreore and Wyatt}(1999)}]{lw1999}
\bibinfo{author}{\bibfnamefont{C.~L.} \bibnamefont{Lopreore}} \bibnamefont{and}
  \bibinfo{author}{\bibfnamefont{R.~E.} \bibnamefont{Wyatt}},
  \bibinfo{journal}{Phys. Rev. Lett} \textbf{\bibinfo{volume}{82}},
  \bibinfo{pages}{5190} (\bibinfo{year}{1999}).

\bibitem[{\citenamefont{Babyuk et~al.}(2003)\citenamefont{Babyuk, Wyatt, and
  Frederick}}]{bwf2003}
\bibinfo{author}{\bibfnamefont{D.}~\bibnamefont{Babyuk}},
  \bibinfo{author}{\bibfnamefont{R.~E.} \bibnamefont{Wyatt}}, \bibnamefont{and}
  \bibinfo{author}{\bibfnamefont{J.~H.} \bibnamefont{Frederick}},
  \bibinfo{journal}{J. Chem. Phys.} \textbf{\bibinfo{volume}{119}},
  \bibinfo{pages}{6482} (\bibinfo{year}{2003}).

\bibitem[{\citenamefont{Beenakker and van Houten}(1991)}]{bvh1991}
\bibinfo{author}{\bibfnamefont{C.~W.} \bibnamefont{Beenakker}}
  \bibnamefont{and} \bibinfo{author}{\bibfnamefont{H.}~\bibnamefont{van
  Houten}}, \bibinfo{journal}{Solid State Phys.} \textbf{\bibinfo{volume}{44}},
  \bibinfo{pages}{1} (\bibinfo{year}{1991}).

\bibitem[{\citenamefont{Wu and Sprung}(1994{\natexlab{a}})}]{ws1994a}
\bibinfo{author}{\bibfnamefont{H.}~\bibnamefont{Wu}} \bibnamefont{and}
  \bibinfo{author}{\bibfnamefont{D.~W.~L.} \bibnamefont{Sprung}},
  \bibinfo{journal}{Phys. Rev. A} \textbf{\bibinfo{volume}{49}},
  \bibinfo{pages}{4305} (\bibinfo{year}{1994}{\natexlab{a}}).

\bibitem[{\citenamefont{Berggren et~al.}(2001)\citenamefont{Berggren, Sadreev,
  and Starikov}}]{bss2001}
\bibinfo{author}{\bibfnamefont{K.~F.} \bibnamefont{Berggren}},
  \bibinfo{author}{\bibfnamefont{A.~F.} \bibnamefont{Sadreev}},
  \bibnamefont{and} \bibinfo{author}{\bibfnamefont{A.~A.}
  \bibnamefont{Starikov}}, \bibinfo{journal}{Nanotechnology}
  \textbf{\bibinfo{volume}{12}}, \bibinfo{pages}{562} (\bibinfo{year}{2001}).

\bibitem[{\citenamefont{Feynman}(1955)}]{feynman1955}
\bibinfo{author}{\bibfnamefont{R.~P.} \bibnamefont{Feynman}},
  \bibinfo{journal}{Progr. Low. Temp. Phys.} \textbf{\bibinfo{volume}{1}},
  \bibinfo{pages}{17} (\bibinfo{year}{1955}).

\bibitem[{\citenamefont{Dalfolo and Stringari}(1996)}]{ds1996}
\bibinfo{author}{\bibfnamefont{F.}~\bibnamefont{Dalfolo}} \bibnamefont{and}
  \bibinfo{author}{\bibfnamefont{S.}~\bibnamefont{Stringari}},
  \bibinfo{journal}{Phys. Rev. A} \textbf{\bibinfo{volume}{53}},
  \bibinfo{pages}{2477} (\bibinfo{year}{1996}).

\bibitem[{\citenamefont{Rokshar}(1997)}]{rok1997}
\bibinfo{author}{\bibfnamefont{D.~S.} \bibnamefont{Rokshar}},
  \bibinfo{journal}{Phys. Rev. Lett.} \textbf{\bibinfo{volume}{79}},
  \bibinfo{pages}{2164} (\bibinfo{year}{1997}).

\bibitem[{\citenamefont{Svidzinsky and Fetter}(1998)}]{sf1998}
\bibinfo{author}{\bibfnamefont{A.~A.} \bibnamefont{Svidzinsky}}
  \bibnamefont{and} \bibinfo{author}{\bibfnamefont{A.~L.}
  \bibnamefont{Fetter}}, \bibinfo{journal}{Phys. Rev. A}
  \textbf{\bibinfo{volume}{58}}, \bibinfo{pages}{310} (\bibinfo{year}{1998}).

\bibitem[{\citenamefont{Duang and Zhang}(1999)}]{dz1999}
\bibinfo{author}{\bibfnamefont{Y.}~\bibnamefont{Duang}} \bibnamefont{and}
  \bibinfo{author}{\bibfnamefont{H.}~\bibnamefont{Zhang}},
  \bibinfo{journal}{Eur. Phys. J.} \textbf{\bibinfo{volume}{D5}},
  \bibinfo{pages}{47} (\bibinfo{year}{1999}).

\bibitem[{\citenamefont{Garcia-Ripoll and Perez-Garcia}(1999)}]{grpg1999}
\bibinfo{author}{\bibfnamefont{J.~J.} \bibnamefont{Garcia-Ripoll}}
  \bibnamefont{and} \bibinfo{author}{\bibfnamefont{V.~M.}
  \bibnamefont{Perez-Garcia}}, \bibinfo{journal}{Phys. Rev. A}
  \textbf{\bibinfo{volume}{60}}, \bibinfo{pages}{4864} (\bibinfo{year}{1999}).

\bibitem[{\citenamefont{Vignolo et~al.}(2007)\citenamefont{Vignolo, Fazio, and
  Tosi}}]{vft2007}
\bibinfo{author}{\bibfnamefont{P.}~\bibnamefont{Vignolo}},
  \bibinfo{author}{\bibfnamefont{R.}~\bibnamefont{Fazio}}, \bibnamefont{and}
  \bibinfo{author}{\bibfnamefont{M.~P.} \bibnamefont{Tosi}},
  \bibinfo{journal}{Phys. Rev. A} \textbf{\bibinfo{volume}{76}},
  \bibinfo{pages}{023616} (\bibinfo{year}{2007}).

\bibitem[{\citenamefont{Sanz et~al.}(2004)\citenamefont{Sanz, Borondo, and
  Miret-Art\'{e}s}}]{sbm2004}
\bibinfo{author}{\bibfnamefont{A.~S.} \bibnamefont{Sanz}},
  \bibinfo{author}{\bibfnamefont{F.}~\bibnamefont{Borondo}}, \bibnamefont{and}
  \bibinfo{author}{\bibnamefont{Miret-Art\'{e}s}}, \bibinfo{journal}{J. Chem.
  Phys.} \textbf{\bibinfo{volume}{120}}, \bibinfo{pages}{8794}
  (\bibinfo{year}{2004}).

\bibitem[{\citenamefont{Bruder et~al.}(1999)\citenamefont{Bruder, Glazman,
  Larkin, Mooij, and can Oudenaarden}}]{bglmo1999}
\bibinfo{author}{\bibfnamefont{C.}~\bibnamefont{Bruder}},
  \bibinfo{author}{\bibfnamefont{L.~I.} \bibnamefont{Glazman}},
  \bibinfo{author}{\bibfnamefont{A.~I.} \bibnamefont{Larkin}},
  \bibinfo{author}{\bibfnamefont{J.~E.} \bibnamefont{Mooij}}, \bibnamefont{and}
  \bibinfo{author}{\bibfnamefont{A.}~\bibnamefont{can Oudenaarden}},
  \bibinfo{journal}{Phys. Rev. B} \textbf{\bibinfo{volume}{59}},
  \bibinfo{pages}{1383} (\bibinfo{year}{1999}).

\bibitem[{\citenamefont{Na and Wyatt}(2002)}]{nw2002}
\bibinfo{author}{\bibfnamefont{K.}~\bibnamefont{Na}} \bibnamefont{and}
  \bibinfo{author}{\bibfnamefont{R.~E.} \bibnamefont{Wyatt}},
  \bibinfo{journal}{Phys. Lett. A} \textbf{\bibinfo{volume}{306}},
  \bibinfo{pages}{97} (\bibinfo{year}{2002}).

\bibitem[{\citenamefont{Wyatt}(2005)}]{w2005}
\bibinfo{author}{\bibfnamefont{R.~E.} \bibnamefont{Wyatt}},
  \emph{\bibinfo{title}{Quantum Dynamics with Trajectories: Introduction to
  Quantum Hydrodynamics}} (\bibinfo{publisher}{Springer}, \bibinfo{address}{New
  York}, \bibinfo{year}{2005}).

\bibitem[{\citenamefont{Madelung}(1926)}]{madelung1926}
\bibinfo{author}{\bibfnamefont{E.}~\bibnamefont{Madelung}},
  \bibinfo{journal}{Z. Phys} \textbf{\bibinfo{volume}{40}},
  \bibinfo{pages}{332} (\bibinfo{year}{1926}).

\bibitem[{\citenamefont{de~Broglie}(1926)}]{brog1926}
\bibinfo{author}{\bibfnamefont{L.}~\bibnamefont{de~Broglie}},
  \bibinfo{journal}{Nature} \textbf{\bibinfo{volume}{118}},
  \bibinfo{pages}{441} (\bibinfo{year}{1926}).

\bibitem[{\citenamefont{Bohm}(1952{\natexlab{a}})}]{bohm1952a}
\bibinfo{author}{\bibfnamefont{D.}~\bibnamefont{Bohm}}, \bibinfo{journal}{Phys.
  Rev.} \textbf{\bibinfo{volume}{85}}, \bibinfo{pages}{166}
  (\bibinfo{year}{1952}{\natexlab{a}}).

\bibitem[{\citenamefont{Bohm}(1952{\natexlab{b}})}]{bohm1952b}
\bibinfo{author}{\bibfnamefont{D.}~\bibnamefont{Bohm}}, \bibinfo{journal}{Phys.
  Rev.} \textbf{\bibinfo{volume}{85}}, \bibinfo{pages}{180}
  (\bibinfo{year}{1952}{\natexlab{b}}).

\bibitem[{\citenamefont{Holland}(2005)}]{hol2005}
\bibinfo{author}{\bibfnamefont{P.}~\bibnamefont{Holland}},
  \bibinfo{journal}{Annals of Physics} \textbf{\bibinfo{volume}{315}},
  \bibinfo{pages}{505} (\bibinfo{year}{2005}).

\bibitem[{\citenamefont{Durr et~al.}(1992)\citenamefont{Durr, Goldstein, and
  Zanghi}}]{dgz1992}
\bibinfo{author}{\bibfnamefont{D.}~\bibnamefont{Durr}},
  \bibinfo{author}{\bibfnamefont{S.}~\bibnamefont{Goldstein}},
  \bibnamefont{and} \bibinfo{author}{\bibfnamefont{N.}~\bibnamefont{Zanghi}},
  \bibinfo{journal}{J. Stat. Phys.} \textbf{\bibinfo{volume}{68}},
  \bibinfo{pages}{259} (\bibinfo{year}{1992}).

\bibitem[{\citenamefont{Faisal and Schwengelbeck}(1995)}]{fs1995}
\bibinfo{author}{\bibfnamefont{F.~H.~M.} \bibnamefont{Faisal}}
  \bibnamefont{and}
  \bibinfo{author}{\bibfnamefont{U.}~\bibnamefont{Schwengelbeck}},
  \bibinfo{journal}{Phys. Lett. A} \textbf{\bibinfo{volume}{207}},
  \bibinfo{pages}{31} (\bibinfo{year}{1995}).

\bibitem[{\citenamefont{Parmenter and Valentine}(1995)}]{pv1995}
\bibinfo{author}{\bibfnamefont{R.~B.} \bibnamefont{Parmenter}}
  \bibnamefont{and} \bibinfo{author}{\bibfnamefont{R.~W.}
  \bibnamefont{Valentine}}, \bibinfo{journal}{Phys. Lett. A}
  \textbf{\bibinfo{volume}{201}}, \bibinfo{pages}{1} (\bibinfo{year}{1995}).

\bibitem[{\citenamefont{de~Polavieja}(1996)}]{pol1996}
\bibinfo{author}{\bibfnamefont{G.~G.} \bibnamefont{de~Polavieja}},
  \bibinfo{journal}{Phys. Rev. A} \textbf{\bibinfo{volume}{53}},
  \bibinfo{pages}{2059} (\bibinfo{year}{1996}).

\bibitem[{\citenamefont{Dewdney and Malik}(1996)}]{dm1996}
\bibinfo{author}{\bibfnamefont{C.}~\bibnamefont{Dewdney}} \bibnamefont{and}
  \bibinfo{author}{\bibfnamefont{Z.}~\bibnamefont{Malik}},
  \bibinfo{journal}{Phys. Lett. A} \textbf{\bibinfo{volume}{220}},
  \bibinfo{pages}{183} (\bibinfo{year}{1996}).

\bibitem[{\citenamefont{Iacomelli and Pettini}(1996)}]{ip1996}
\bibinfo{author}{\bibfnamefont{G.}~\bibnamefont{Iacomelli}} \bibnamefont{and}
  \bibinfo{author}{\bibfnamefont{M.}~\bibnamefont{Pettini}},
  \bibinfo{journal}{Phys. Lett. A} \textbf{\bibinfo{volume}{212}},
  \bibinfo{pages}{29} (\bibinfo{year}{1996}).

\bibitem[{\citenamefont{Frisk}(1997)}]{frisk1997}
\bibinfo{author}{\bibfnamefont{H.}~\bibnamefont{Frisk}},
  \bibinfo{journal}{Phys. Lett. A} \textbf{\bibinfo{volume}{227}},
  \bibinfo{pages}{139} (\bibinfo{year}{1997}).

\bibitem[{\citenamefont{Konkel and Makowski}(1998)}]{km1998}
\bibinfo{author}{\bibfnamefont{S.}~\bibnamefont{Konkel}} \bibnamefont{and}
  \bibinfo{author}{\bibfnamefont{A.~J.} \bibnamefont{Makowski}},
  \bibinfo{journal}{Phys. Lett. A} \textbf{\bibinfo{volume}{238}},
  \bibinfo{pages}{95} (\bibinfo{year}{1998}).

\bibitem[{\citenamefont{Wu and Sprung}(1999)}]{ws1999}
\bibinfo{author}{\bibfnamefont{H.}~\bibnamefont{Wu}} \bibnamefont{and}
  \bibinfo{author}{\bibfnamefont{D.~W.~L.} \bibnamefont{Sprung}},
  \bibinfo{journal}{Phys. Lett. A} \textbf{\bibinfo{volume}{261}},
  \bibinfo{pages}{150} (\bibinfo{year}{1999}).

\bibitem[{\citenamefont{Makowski et~al.}(2000)\citenamefont{Makowski,
  Peplowski, and Dembinski}}]{mpd2000}
\bibinfo{author}{\bibfnamefont{A.~J.} \bibnamefont{Makowski}},
  \bibinfo{author}{\bibfnamefont{P.}~\bibnamefont{Peplowski}},
  \bibnamefont{and} \bibinfo{author}{\bibfnamefont{S.~T.}
  \bibnamefont{Dembinski}}, \bibinfo{journal}{Phys. Lett. A}
  \textbf{\bibinfo{volume}{266}}, \bibinfo{pages}{241} (\bibinfo{year}{2000}).

\bibitem[{\citenamefont{Cushing}(2000)}]{cush2000}
\bibinfo{author}{\bibfnamefont{J.~T.} \bibnamefont{Cushing}},
  \bibinfo{journal}{Philosophy of Science} \textbf{\bibinfo{volume}{67}},
  \bibinfo{pages}{S432} (\bibinfo{year}{2000}).

\bibitem[{\citenamefont{Falsaperla and Fonte}(2003)}]{ff2003}
\bibinfo{author}{\bibfnamefont{P.}~\bibnamefont{Falsaperla}} \bibnamefont{and}
  \bibinfo{author}{\bibfnamefont{G.}~\bibnamefont{Fonte}},
  \bibinfo{journal}{Phys. Lett. A} \textbf{\bibinfo{volume}{316}},
  \bibinfo{pages}{382} (\bibinfo{year}{2003}).

\bibitem[{\citenamefont{de~Sales and Florencio}(2003)}]{sf2003}
\bibinfo{author}{\bibfnamefont{J.~A.} \bibnamefont{de~Sales}} \bibnamefont{and}
  \bibinfo{author}{\bibfnamefont{J.}~\bibnamefont{Florencio}},
  \bibinfo{journal}{Phys. Rev. E} \textbf{\bibinfo{volume}{67}},
  \bibinfo{pages}{016216} (\bibinfo{year}{2003}).

\bibitem[{\citenamefont{Wisniacki and Pujals}(2005)}]{wp2005}
\bibinfo{author}{\bibfnamefont{D.~A.} \bibnamefont{Wisniacki}}
  \bibnamefont{and} \bibinfo{author}{\bibfnamefont{E.~R.}
  \bibnamefont{Pujals}}, \bibinfo{journal}{Europhys. Lett.}
  \textbf{\bibinfo{volume}{71}}, \bibinfo{pages}{159} (\bibinfo{year}{2005}).

\bibitem[{\citenamefont{Valentini and Westman}(2005)}]{vw2005}
\bibinfo{author}{\bibfnamefont{A.}~\bibnamefont{Valentini}} \bibnamefont{and}
  \bibinfo{author}{\bibfnamefont{H.}~\bibnamefont{Westman}},
  \bibinfo{journal}{Proc. R. Soc. A} \textbf{\bibinfo{volume}{461}},
  \bibinfo{pages}{253} (\bibinfo{year}{2005}).

\bibitem[{\citenamefont{Wisniacki et~al.}(2007)\citenamefont{Wisniacki, Pujals,
  and Borondo}}]{wpb2007}
\bibinfo{author}{\bibfnamefont{D.~A.} \bibnamefont{Wisniacki}},
  \bibinfo{author}{\bibfnamefont{E.~R.} \bibnamefont{Pujals}},
  \bibnamefont{and} \bibinfo{author}{\bibfnamefont{F.}~\bibnamefont{Borondo}},
  \bibinfo{journal}{J. Phys. A} \textbf{\bibinfo{volume}{40}},
  \bibinfo{pages}{14353} (\bibinfo{year}{2007}).

\bibitem[{\citenamefont{Schlegel and Forster}(2008)}]{sf2008}
\bibinfo{author}{\bibfnamefont{K.~G.} \bibnamefont{Schlegel}} \bibnamefont{and}
  \bibinfo{author}{\bibfnamefont{S.}~\bibnamefont{Forster}},
  \bibinfo{journal}{Phys. Lett. A} \textbf{\bibinfo{volume}{372}},
  \bibinfo{pages}{3620} (\bibinfo{year}{2008}).

\bibitem[{\citenamefont{Efthymiopoulos and Contopoulos}(2006)}]{efthcont2006}
\bibinfo{author}{\bibfnamefont{C.}~\bibnamefont{Efthymiopoulos}}
  \bibnamefont{and}
  \bibinfo{author}{\bibfnamefont{G.}~\bibnamefont{Contopoulos}},
  \bibinfo{journal}{J. Phys. A} \textbf{\bibinfo{volume}{39}},
  \bibinfo{pages}{1819} (\bibinfo{year}{2006}).

\bibitem[{\citenamefont{Efthymiopoulos
  et~al.}(2007)\citenamefont{Efthymiopoulos, Kalapotharakos, and
  Contopoulos}}]{ekc2007}
\bibinfo{author}{\bibfnamefont{C.}~\bibnamefont{Efthymiopoulos}},
  \bibinfo{author}{\bibfnamefont{C.}~\bibnamefont{Kalapotharakos}},
  \bibnamefont{and}
  \bibinfo{author}{\bibfnamefont{G.}~\bibnamefont{Contopoulos}},
  \bibinfo{journal}{J. Phys. A} \textbf{\bibinfo{volume}{40}},
  \bibinfo{pages}{12945} (\bibinfo{year}{2007}).

\bibitem[{\citenamefont{Contopoulos and Efthymiopoulos}(2008)}]{contefth2008}
\bibinfo{author}{\bibfnamefont{G.}~\bibnamefont{Contopoulos}} \bibnamefont{and}
  \bibinfo{author}{\bibfnamefont{C.}~\bibnamefont{Efthymiopoulos}},
  \bibinfo{journal}{Cel. Mech. Dyn. Astron.} \textbf{\bibinfo{volume}{102}},
  \bibinfo{pages}{219} (\bibinfo{year}{2008}).

\bibitem[{\citenamefont{Wu and Sprung}(1994{\natexlab{b}})}]{ws1994b}
\bibinfo{author}{\bibfnamefont{H.}~\bibnamefont{Wu}} \bibnamefont{and}
  \bibinfo{author}{\bibfnamefont{D.~W.~L.} \bibnamefont{Sprung}},
  \bibinfo{journal}{Phys. Lett. A} \textbf{\bibinfo{volume}{196}},
  \bibinfo{pages}{229} (\bibinfo{year}{1994}{\natexlab{b}}).

\bibitem[{\citenamefont{Berry}(2005)}]{berry2005}
\bibinfo{author}{\bibfnamefont{M.~V.} \bibnamefont{Berry}},
  \bibinfo{journal}{J. Phys. A} \textbf{\bibinfo{volume}{38}},
  \bibinfo{pages}{L745} (\bibinfo{year}{2005}).

\bibitem[{\citenamefont{Contopoulos et~al.}(1997)\citenamefont{Contopoulos,
  Voglis, Efthymiopoulos, Froeschl\'{e}, Gonczi, Lega, Dvorak, and
  Lohinger}}]{cve1997}
\bibinfo{author}{\bibfnamefont{G.}~\bibnamefont{Contopoulos}},
  \bibinfo{author}{\bibfnamefont{N.}~\bibnamefont{Voglis}},
  \bibinfo{author}{\bibfnamefont{C.}~\bibnamefont{Efthymiopoulos}},
  \bibinfo{author}{\bibfnamefont{C.}~\bibnamefont{Froeschl\'{e}}},
  \bibinfo{author}{\bibfnamefont{R.}~\bibnamefont{Gonczi}},
  \bibinfo{author}{\bibfnamefont{E.}~\bibnamefont{Lega}},
  \bibinfo{author}{\bibfnamefont{R.}~\bibnamefont{Dvorak}}, \bibnamefont{and}
  \bibinfo{author}{\bibfnamefont{E.}~\bibnamefont{Lohinger}},
  \bibinfo{journal}{Cel. Mech. Dyn. Astron.} \textbf{\bibinfo{volume}{67}},
  \bibinfo{pages}{293} (\bibinfo{year}{1997}).

\bibitem[{\citenamefont{Contopoulos}(2004)}]{cont2002}
\bibinfo{author}{\bibfnamefont{G.}~\bibnamefont{Contopoulos}},
  \emph{\bibinfo{title}{Order and Chaos in Dynamical Astronomy}}
  (\bibinfo{publisher}{Springer-Verlag}, \bibinfo{address}{New York},
  \bibinfo{year}{2004}).

\end{thebibliography}

\end{document}